\documentclass[
aps,
reprint,
prapplied,
superscriptaddress,
10pt
]{revtex4-2}

\usepackage{silence}
\usepackage[T1]{fontenc}
\usepackage{amsmath,amssymb}
\usepackage{multirow,booktabs}
\usepackage{graphicx}
\usepackage{here}
\usepackage[
  dvipdfmx,
  colorlinks=true,
  citecolor=cyan,
  linkcolor=black,
  urlcolor=magenta
]{hyperref}
\usepackage{cleveref}
\usepackage{tikz}
\usepackage{placeins}
\usepackage{chngcntr}

\allowdisplaybreaks[4]

\usetikzlibrary{calc,positioning}

\crefname{equation}{Eq.}{Eqs.}
\Crefname{equation}{Equation}{Equations}
\creflabelformat{equation}{#2(#1)#3}
\crefname{table}{Table}{Tables}
\Crefname{table}{Table}{Tables}
\crefname{figure}{Fig.}{Figs.}
\Crefname{figure}{Figure}{Figures}
\crefname{section}{Sec.}{Secs.}
\Crefname{section}{Section}{Sections}
\crefname{appendix}{Appendix}{Appendices}
\Crefname{appendix}{Appendix}{Appendices}
\AddToHook{cmd/appendix/before}{\crefalias{section}{appendix}}

\begin{document}
\title{Linear optical Bell state measurement for rotation-symmetric cat codes}
\author{Issa Oe}
\email{issa.oe@ntt.com}
\affiliation{NTT Computer and Data Science Laboratories, NTT Inc., Musashino 180-8585, Japan}

\author{Suguru Endo}
\affiliation{NTT Computer and Data Science Laboratories, NTT Inc., Musashino 180-8585, Japan}
\affiliation{NTT Research Center for Theoretical Quantum Information, NTT Inc. 3-1 Morinosato Wakanomiya, Atsugi, Kanagawa, 243-0198, Japan}

\author{Rui Asaoka}
\email{rui.asaoka@ntt.com}
\affiliation{NTT Computer and Data Science Laboratories, NTT Inc., Musashino 180-8585, Japan}
\affiliation{NTT Research Center for Theoretical Quantum Information, NTT Inc. 3-1 Morinosato Wakanomiya, Atsugi, Kanagawa, 243-0198, Japan}

\begin{abstract}
	Rotation-symmetric cat (RS-cat) codes are a bosonic-code platform for quantum information processing, combining finite-energy realizability with robustness against photon loss through their discrete rotational symmetry. For applications in long-distance quantum communication and fusion-based quantum computation (FBQC), efficient Bell state measurement (BSM) is a key primitive. In this work, we consider a BSM protocol for RS-cat codes using only a half beam splitter (HBS) and photon-number-resolving detectors (PNRDs). By exploiting the characteristic photon-number structure induced by the discrete rotational symmetry of RS-cat codes, our protocol extracts both photon-number modulo and phase information for Bell-state discrimination. We show that, under ideal loss-free conditions, the proposed BSM protocol becomes deterministic for arbitrary symmetry order $N$ for sufficiently large amplitudes $\alpha$. We further numerically evaluate the success probability under photon loss and identify the loss regime in which higher-order RS-cat codes provide an advantage. Finally, we show that post-selection can enhance the success probability.
\end{abstract}

\maketitle

\section{Introduction}

% 1. CV & ボソニック符号について
Continuous-variable quantum computing \cite{lloyd1999quantum, braunstein2005quantum} is a promising route toward practical quantum computation, as bosonic modes arise naturally in many platforms and can be controlled with quantum-optical and circuit-QED techniques. Nevertheless, the presence of noise such as photon loss and dephasing necessitates quantum error correction. Bosonic codes, including Gottesman–Kitaev–Preskill (GKP) codes \cite{gottesman2000encoding, grimsmo2021quantum}, cat codes \cite{cochrane1999macroscopically, mirrahimi2014dynamically, ofek2016extending}, and binomial codes \cite{michael2016new}, address this challenge by encoding logical information into structured oscillator states, offering a hardware-efficient path toward fault-tolerant quantum computation.

% 2. GKP 符号の無限エネルギー性
GKP codes, often regarded as a leading candidate for fault-tolerant quantum computation, encode quantum information into grid states forming a periodic lattice in phase space \cite{gottesman2000encoding}. This structure enables, in principle, the correction of small displacement errors \cite{glancy2006error}. However, ideal GKP codewords require infinitely squeezed, infinite-energy states and are therefore unphysical. Realistic finite-energy GKP states inevitably suffer from envelope-induced distortions and effective displacement errors, which degrade state preparation, gates, and measurement \cite{glancy2006error, fukui2018high, matsuura2020equivalence, grimsmo2021quantum, rojkov2024two, jafarzadeh2025logical, matsuura2026continuous}. Thus, while GKP codes provide a powerful framework for displacement-noise correction, their performance is strongly constrained by finite-energy effects.

% 3. RS-cat code の有限エネルギー性
In contrast, rotation-symmetric cat (RS-cat) codes \cite{grimsmo2020quantum, hillmann2022performance, marinoff2024explicit, li2024performance, endo2025quantum, udupa2025performance, my2025circuit} are built from finite-energy superpositions of coherent states with discrete rotational symmetry in phase space. Rather than relying on an ideal infinite-energy lattice, RS-cat codes exploit the modular photon-number structure induced by their discrete rotational symmetry. Photon loss then acts as a discrete shift in photon-number parity, allowing an $N$-fold RS-cat code to detect loss errors up to order $N-1$. The conventional cat code is recovered as the twofold-symmetric case. This makes RS-cat codes a natural bosonic-code platform for studying loss-robust operations under realistic finite-energy conditions.

% 4. RS-cat code を BSMすることの重要性について
For applying RS-cat codes to long-distance quantum communication \cite{fukui2021all, rozpkedek2021quantum, rozpkedek2023all, azari2024quantum} and fusion-based quantum computation (FBQC) \cite{browne2005resource, bartolucci2023fusion}, efficient Bell state measurement (BSM) \cite{jozsateleporting, zukowski1993event, braunstein1995measurement, lutkenhaus1999bell} is a key primitive. BSM connects independently prepared entangled resources, enabling teleportation \cite{bartlett2003quantum, zhou2000methodology}, entanglement swapping \cite{zukowski1993event, pan1998experimental}, and fusion operations \cite{browne2005resource, bartolucci2023fusion}. Its efficiency and loss robustness therefore directly determine the performance and scalability of bosonic-code-based communication and computation architectures.

% 5. BSM の難しさについて
Nevertheless, designing such BSM protocols for bosonic codes is challenging. This difficulty arises from the infinite-dimensional Hilbert space of bosonic modes, the non-orthogonality of the code states, and the complicated effects of photon loss. This naturally raises the question of what can be achieved in bosonic-code-based BSM under minimal optical and measurement resources, and how the structure of the code influences Bell state distinguishability. While BSM protocols for GKP codes have been well studied \cite{fukui2021all, schmidt2024error}, BSM protocols for rotation-symmetric bosonic codes, including RS-cat codes, remain comparatively less explored.

\begin{figure}[htbp]
	\centering
	\includegraphics[width=0.7\linewidth]{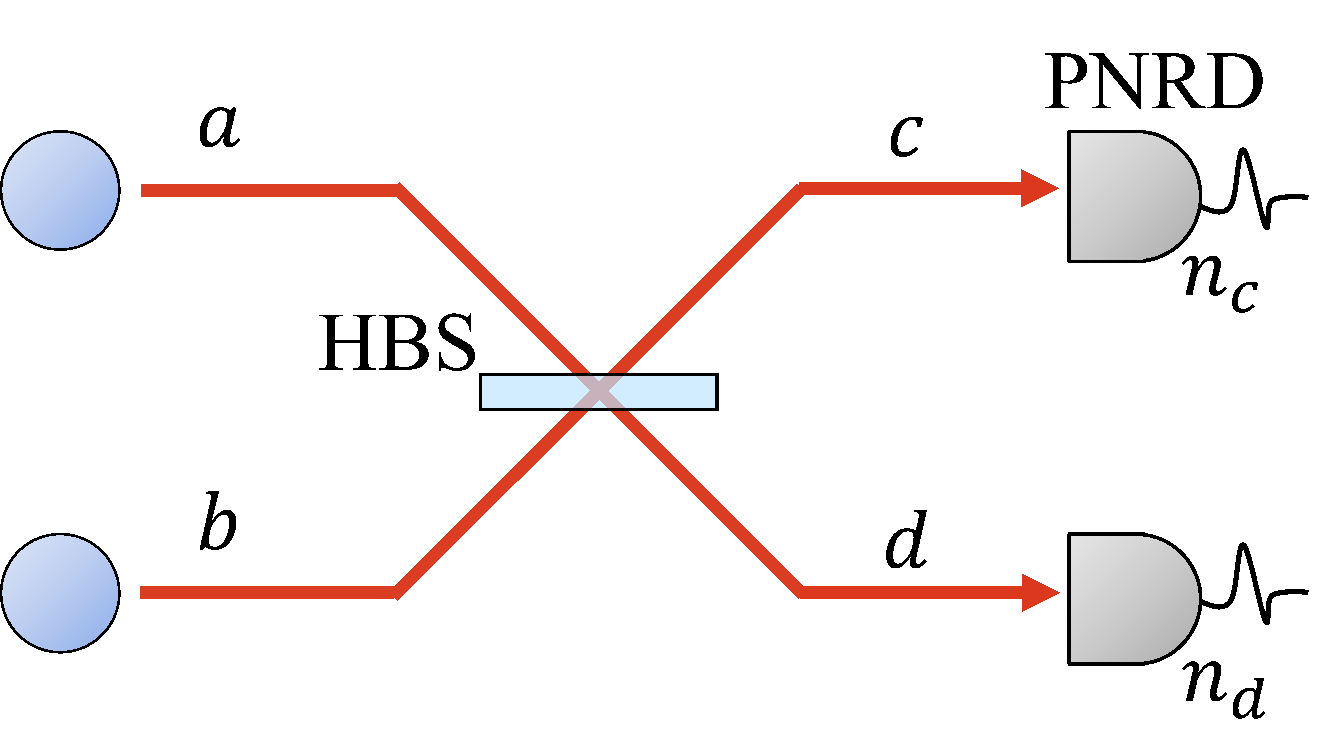}
	\caption{Schematic of the BSM setup consisting of a HBS and two PNRDs.}
	\label{fig:optical_device}
\end{figure}

% 6. 手法の詳細
Motivated by this question, we consider BSM for RS-cat codes with arbitrary symmetry order using only a half beam splitter (HBS) and photon-number-resolving detectors (PNRDs), which provides a more experimentally accessible route to CV quantum information processing based on RS-cat codes. A previous approach by Grimsmo \textit{et al.} \cite{grimsmo2020quantum} employs nonlinear operations such as controlled-rotations (CROT) in circuit-QED implementations of BSMs, making the scheme platform-dependent. In contrast, here we consider a setup based only on minimal bosonic operations, as shown in \cref{fig:optical_device}.

% 7. 成果
Our strategy is based on extracting both photon-number modulo information and phase information, which together enable the identification of the Bell states. We show that, for sufficiently large amplitude $\alpha$, the protocol becomes effectively deterministic for arbitrary symmetry order $N$ in the absence of photon loss. We then numerically investigate its robustness against photon loss by evaluating the success probability as a function of the loss rate, and identify the regime in which higher-order RS-cat codes provide an advantage. We further demonstrate that post-selection can enhance the conditional success probability. Finally, we analyze the photon-number resolution required by our BSM protocol and show that it remains within a potentially achievable range.

% 8. 論文構成
This paper is structured as follows. In \cref{sec:preliminary}, after reviewing previous BSM methods, we introduce the definition and characteristics of RS-cat codes. \Cref{sec:bsm_rs_cat_1,sec:bsm_rs_cat_2} theoretically analyzes how to perform BSM on RS-cat codes. \Cref{sec:without_loss} presents the numerical simulation results of the BSM, and \cref{sec:with_loss} conducts an analysis including photon loss to investigate the relationship between the symmetry order and error tolerance. Additionally, \cref{sec:post-selection} demonstrates a method to improve the success rate via post-selection. Finally, \cref{sec:conclusion} concludes the paper and discusses future directions.

\section{preliminaries}
\label{sec:preliminary}

In this section, we first review BSM schemes for the coherent code and the cat code, which serve as useful references for developing a BSM scheme for RS-cat codes. We then introduce the RS-cat code framework used throughout this work. BSM is the task of discriminating the following four logical Bell states:
\begin{equation}
	\begin{aligned}
		\label{eq:bell_state}
		|\Phi^\pm_L\rangle
		 & = \frac{1}{\sqrt{2}}\left(|0_L0_L\rangle \pm |1_L1_L\rangle\right), \\
		|\Psi^\pm_L\rangle
		 & = \frac{1}{\sqrt{2}}\left(|0_L1_L\rangle \pm |1_L0_L\rangle\right),
	\end{aligned}
\end{equation}
where $|0_L\rangle$ and $|1_L\rangle$ are the logical basis states. In both the coherent code and the cat code, BSM is implemented by interfering the two modes on a HBS and measuring the output modes with PNRDs. The Bell state is then identified from the detected photon-number pattern $(n_c, n_d)$. We adopt the convention in which the HBS unitary operator $U_{\mathrm{HBS}}$ transforms the input-mode annihilation operators $a, b$ into the output-mode annihilation operators $c,d$ as
\begin{equation}
	a \mapsto \frac{c + d}{\sqrt2}, \qquad
	b \mapsto \frac{c - d}{\sqrt2}.
\end{equation}
When two coherent states $|\alpha\rangle, |\beta\rangle \; (\alpha, \beta \in \mathbb{C})$ are input to the HBS, they are transformed as follows:
\begin{equation}
	\label{eq:bs_coherent}
	U_{\mathrm{HBS}}|\alpha\rangle_a |\beta\rangle_b = \left| \frac{\alpha + \beta}{\sqrt{2}} \right\rangle_c \left| \frac{\alpha - \beta}{\sqrt{2}} \right\rangle_d.
\end{equation}

\subsection{BSM for the coherent code}

The logical basis of the coherent code is defined using the coherent state $|\alpha\rangle$ with amplitude $\alpha \in \mathbb{R}$ as
\begin{equation}
	\begin{aligned}
		|0_L\rangle & = |\alpha\rangle,  \\
		|1_L\rangle & = |-\alpha\rangle.
	\end{aligned}
\end{equation}
Applying \cref{eq:bs_coherent} to the Bell states yields
\begin{equation}
	\begin{aligned}
		U_{\mathrm{HBS}} |\Phi^+_L\rangle & = |\mathrm{Cat}_{\alpha'}^+\rangle_c \; |0\rangle_d, \\
		U_{\mathrm{HBS}} |\Phi^-_L\rangle & = |\mathrm{Cat}_{\alpha'}^-\rangle_c \; |0\rangle_d, \\
		U_{\mathrm{HBS}} |\Psi^+_L\rangle & = |0\rangle_c \; |\mathrm{Cat}_{\alpha'}^+\rangle_d, \\
		U_{\mathrm{HBS}} |\Psi^-_L\rangle & = |0\rangle_c \; |\mathrm{Cat}_{\alpha'}^-\rangle_d,
	\end{aligned}
\end{equation}
where
\begin{equation}
	\alpha' = \sqrt2\alpha,
	\quad
	|\mathrm{Cat}_{\alpha'}^\pm\rangle
	= \mathcal{N}_\pm \left(|\alpha'\rangle \pm |-\alpha'\rangle\right),
\end{equation}
and $\mathcal{N}_\pm \in \mathbb{R}$ are normalization constants. Since $|\mathrm{Cat}_{\alpha'}^+\rangle$ and $|\mathrm{Cat}_{\alpha'}^-\rangle$ contain only even and odd photon-number components, respectively, the Bell states can be identified by measuring the photon-number pattern of the output modes, as summarized in \cref{table:bsm_coherent_code}.

\begin{table}[htbp]
	\centering
	\caption{Decision rules for Bell-state identification with the coherent code.}
	\begin{tabular}{ccccccc}
		\toprule
		\multicolumn{5}{c}{Measurement Pattern}
		 & Decision                                       \\
		\cmidrule(r){1-5} \cmidrule(l){6-6}
		\quad\quad
		 & \multirow{2}{*}{$n_c + n_d \equiv 0 \pmod{2}$}
		 & \quad\quad\quad
		 & $n_d = 0$
		 & \quad\quad
		 & $|\Phi^+_L\rangle$
		 & \quad                                          \\
		 &
		 &
		 & $n_c = 0$
		 &
		 & $|\Psi^+_L\rangle$                             \\
		\midrule
		 & \multirow{2}{*}{$n_c + n_d \equiv 1 \pmod{2}$}
		 &
		 & $n_d = 0$
		 &
		 & $|\Phi^-_L\rangle$                             \\
		 &
		 &
		 & $n_c = 0$
		 &
		 & $|\Psi^-_L\rangle$                             \\
		\bottomrule
	\end{tabular}
	\label{table:bsm_coherent_code}
\end{table}

\noindent
In the absence of photon loss, the only intrinsic failure event occurs when no photons are detected at either output port, in which case $|\Phi^+_L\rangle$ and $|\Psi^+_L\rangle$ remain indistinguishable. As amplitude $\alpha$ increases, the probability of this no-click event becomes negligible, allowing the BSM to approach deterministic operation in the ideal limit. Photon loss, however, introduces an additional failure mechanism: a single-photon loss event flips the photon-number parity, thereby leading to an incorrect BSM outcome.

\subsection{BSM for the cat code}

The cat code is defined using coherent states $|\alpha\rangle$ with amplitude $\alpha \in \mathbb{R}$ as
\begin{equation}
	\begin{aligned}
		|0_L\rangle & = \mathcal{N}_+ \left(|\alpha\rangle + |-\alpha\rangle\right),   \\
		|1_L\rangle & = \mathcal{N}_+ \left(|i\alpha\rangle + |-i\alpha\rangle\right).
	\end{aligned}
\end{equation}
When the Bell states of the cat code are input into a HBS, the output states are given by
\begin{equation}
	\begin{aligned}
		U_{\mathrm{HBS}} |\Phi^+_L\rangle
		 & = \frac{1}{\sqrt{2}}\left(
		|C_0^{(\alpha')}\rangle_c |0\rangle_d
		+ |0\rangle_c |C_0^{(\alpha')}\rangle_d
		\right),                       \\
		U_{\mathrm{HBS}} |\Phi^-_L\rangle
		 & = \frac{1}{\sqrt{2}}\left(
		|C_2^{(\alpha')}\rangle_c |0\rangle_d
		+ |0\rangle_c |C_2^{(\alpha')}\rangle_d
		\right),                       \\
		U_{\mathrm{HBS}} |\Psi^+_L\rangle
		 & = \frac{1}{\sqrt{2}} \left(
		|\mathrm{Cat}^+_{\tilde{\alpha}}\rangle_c
		|\mathrm{Cat}^+_{\tilde{\alpha}^*}\rangle_d
		+ |\mathrm{Cat}^+_{\tilde{\alpha}^*}\rangle_c
		|\mathrm{Cat}^+_{\tilde{\alpha}}\rangle_d
		\right),                       \\
		U_{\mathrm{HBS}} |\Psi^-_L\rangle
		 & = \frac{1}{\sqrt{2}} \left(
		|\mathrm{Cat}^-_{\tilde{\alpha}}\rangle_c
		|\mathrm{Cat}^-_{\tilde{\alpha}^*}\rangle_d
		+ |\mathrm{Cat}^-_{\tilde{\alpha}^*}\rangle_c
		|\mathrm{Cat}^-_{\tilde{\alpha}}\rangle_d
		\right),
	\end{aligned}
\end{equation}
where
\begin{equation}
	\begin{aligned}
		 & \tilde{\alpha} = \tfrac{1 + i}{\sqrt2}\alpha, \quad
		\tilde{\alpha}^* = \tfrac{1 - i}{\sqrt2}\alpha,        \\
		 & |C_0^{(\alpha')}\rangle
		= \mathcal{M}_0 \left(
		|\alpha'\rangle
		+ |-\alpha'\rangle
		+ |i\alpha'\rangle
		+ |-i\alpha'\rangle
		\right),                                               \\
		 & |C_2^{(\alpha')}\rangle
		= \mathcal{M}_2 \left(
		|\alpha'\rangle
		+ |-\alpha'\rangle
		- |i\alpha'\rangle
		- |-i\alpha'\rangle
		\right),
	\end{aligned}
\end{equation}
and $\mathcal{M}_0, \mathcal{M}_2 \in \mathbb{R}$ are normalization constants. Since $|C_0^{(\alpha')}\rangle$ and $|C_2^{(\alpha')}\rangle$ consist only of $4k$ and $4k+2$ photon numbers, respectively, the BSM for the cat code proposed by Hastrup \textit{et al.} \cite{hastrup2022all} identifies the Bell states according to \cref{table:bsm_cat_code}.

\begin{table}[htbp]
	\centering
	\caption{Decision rules for Bell-state identification with the cat code.}
	\begin{tabular}{ccccccc}
		\toprule
		\multicolumn{5}{c}{Measurement Pattern}
		 & Decision                                       \\
		\cmidrule(r){1-5} \cmidrule(l){6-6}
		\quad\quad
		 & \multirow{2}{*}{$n_c + n_d \equiv 0 \pmod{4}$}
		 & \quad\quad
		 & $n_c = 0$ or $n_d = 0$
		 & \quad
		 & $|\Phi^+_L\rangle$
		 & \quad                                          \\
		 &
		 &
		 & otherwise
		 &
		 & $|\Psi^+_L\rangle$                             \\
		\midrule
		 & \multirow{2}{*}{$n_c + n_d \equiv 2 \pmod{4}$}
		 &
		 & $n_c = 0$ or $n_d = 0$
		 &
		 & $|\Phi^-_L\rangle$                             \\
		 &
		 &
		 & otherwise
		 &
		 & $|\Psi^-_L\rangle$                             \\
		\bottomrule
	\end{tabular}
	\label{table:bsm_cat_code}
\end{table}

\noindent
In the lossless case, the only intrinsic failure of this BSM occurs for $|\Psi^\pm_L\rangle$ when either output mode yields a zero-photon outcome. Since this pattern is assigned to $|\Phi^\pm_L\rangle$, such $|\Psi^\pm_L\rangle$ events are misidentified as $|\Phi^\pm_L\rangle$. This failure probability becomes negligible for large $\alpha$, and the BSM approaches deterministic operation in the ideal limit. With photon loss, however, additional errors arise, since lost photons can alter the detected photon-number pattern and lead to an incorrect Bell-state identification.

Around the same time, Su \textit{et al.} \cite{su2022universal} proposed another BSM method for the cat code. Their scheme utilizes three HBSs and four PNRDs. However, as noted by Lee \textit{et al.} \cite{lee2024fault}, the BSM scheme proposed by Hastrup \textit{et al.} achieves a higher success probability than Su \textit{et al.}'s scheme.

\begin{figure*}[tb]
	\begin{minipage}{0.2\textwidth}
		\centering
		\small{(coherent)$N=1,\alpha=2.0$}
		\includegraphics[width=\textwidth]{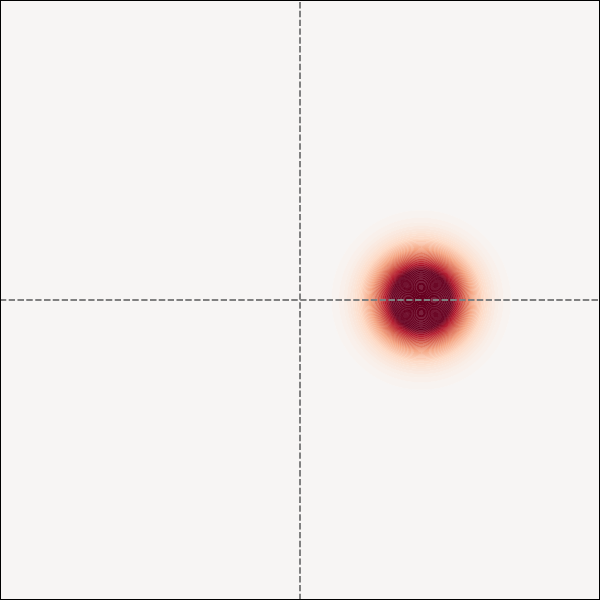}
	\end{minipage}
	\hspace{10pt}
	\begin{minipage}{0.2\textwidth}
		\centering
		\small{(cat) $N=2,\alpha=2.0$}
		\includegraphics[width=\textwidth]{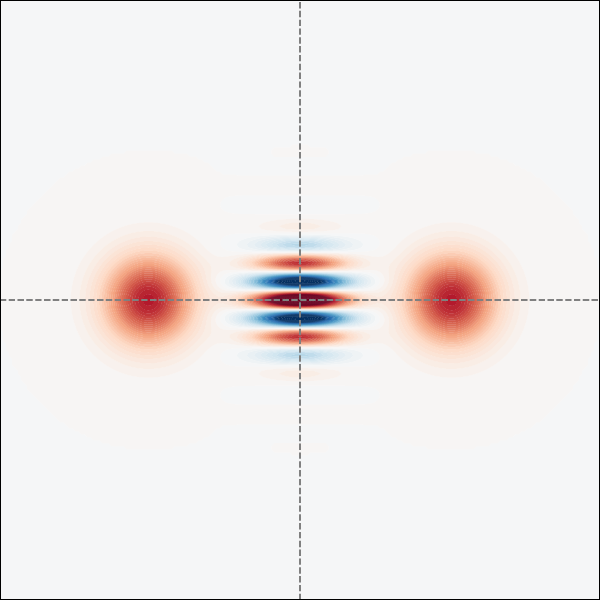}
	\end{minipage}
	\hspace{10pt}
	\begin{minipage}{0.2\textwidth}
		\centering
		\small{$N=3,\alpha=3.5$}
		\includegraphics[width=\textwidth]{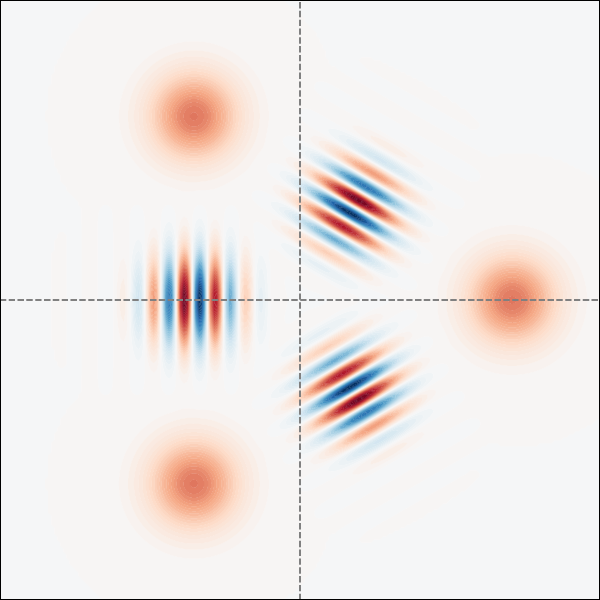}
	\end{minipage}
	\hspace{10pt}
	\begin{minipage}{0.2\textwidth}
		\centering
		\small{$N=4,\alpha=4.0$}
		\includegraphics[width=\textwidth]{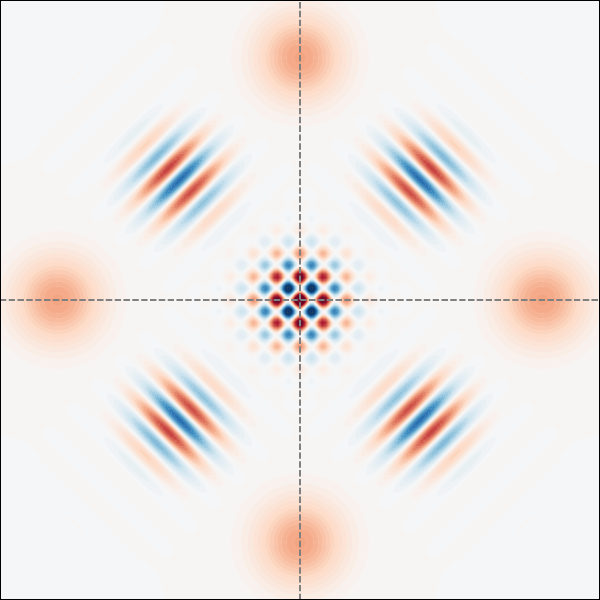}
	\end{minipage}
	\caption{Wigner functions of RS-cat codes. The panels show the cases $N=1,2,3,\textrm{ and }4$, with the corresponding values of $\alpha$ indicated above each plot.}
	\label{fig:wigner_function}
\end{figure*}

\subsection{Definition and characteristics of RS-cat codes}
\label{sec:rs_cat}
We now introduce the definition and characteristics of RS-cat codes. RS-cat codes are bosonic codes that extend the rotational symmetry of the cat code in phase space. Let $\alpha \in \mathbb{R}$ be the amplitude and $N \in \mathbb{N}$ be the symmetry order. The logical basis states of RS-cat codes are defined as superpositions of coherent states:
\begin{equation}
	\begin{aligned}
		\label{eq:def_rs_cat}
		|0_L\rangle
		 & = \mathcal{N}_{\mathrm{RS}}^{(N)} \sum_{m=0}^{N-1} |e^{\frac{2m\pi i}{N}}\alpha\rangle,     \\
		|1_L\rangle
		 & = \mathcal{N}_{\mathrm{RS}}^{(N)} \sum_{m=0}^{N-1} |e^{\frac{(2m+1)\pi i}{N}}\alpha\rangle,
	\end{aligned}
\end{equation}
where $\mathcal{N}_{\mathrm{RS}}^{(N)} \in \mathbb{R}$ is a normalization constant. A coherent state $|\alpha\rangle$ can be expanded in the Fock basis as
\begin{equation}
	|\alpha\rangle = e^{-\frac{|\alpha|^2}{2}} \sum_{n=0}^{\infty} \frac{\alpha^n}{\sqrt{n!}} |n\rangle,
\end{equation}
where the photon number is distributed according to a Poisson distribution with mean $|\alpha|^2$:
\begin{equation}
	\label{eq:poisson_distribution}
	n  \sim \mathrm{Poisson}(|\alpha|^2).
\end{equation}
Using this expansion, the logical codewords can be expanded in the Fock basis as
\begin{equation}
	\label{eq:fock_rs_cat}
	|0_L\rangle = \sum_{k=0}^{\infty} f_k |kN\rangle, \quad
	|1_L\rangle = \sum_{k=0}^{\infty} (-1)^k f_k |kN\rangle,
\end{equation}
where
\begin{equation}
	f_k = \frac{\mathcal{C}_{N,\alpha} \; \alpha^{kN}}{\sqrt{(kN)!}},
\end{equation}
with $\mathcal{C}_{N,\alpha} \in \mathbb{R}$ being a normalization constant. Thus, an $N$-fold RS-cat code consists only of Fock states whose photon numbers are multiples of $N$. On the other hand, when photon loss occurs, the state includes photon-number components that are no longer multiples of $N$. By detecting such changes, the occurrence of an error can be identified. Consequently, increasing the symmetry order $N$ provides robustness not only against single-photon loss but also against higher-order photon-loss events. Note that the coherent code and the cat code are recovered as the $N = 1$ and $N = 2$ cases of the RS-cat code, respectively. The Wigner functions of the RS-cat codes for $N = 1, 2, 3,$ and $4$ are shown in \cref{fig:wigner_function}.

\section{BSM for general symmetry order RS-cat codes}

Here we describe our BSM protocol for RS-cat codes. Similar to the BSM for the coherent code and the cat code, our BSM protocol uses the setup in \cref{fig:optical_device}. \Cref{table:bsm_rs_cat_without_loss} summarizes the decision rules for identifying the Bell states from the detected photon-number patterns $(n_c, n_d)$.

\begin{table}[H]
	\centering
	\caption{Decision rules for Bell-state identification with RS-cat codes.}
	\begin{tabular}{ccc}
		\toprule
		\multicolumn{2}{c}{Measurement Pattern}
		 & Decision                        \\
		\cmidrule(r){1-2} \cmidrule(l){3-3}
		\multirow{2}{*}{$n_c + n_d \equiv 0 \pmod{2N}$}
		 & $D(n_c, n_d) \equiv 0 \pmod{2}$
		 & $|\Phi^+_L\rangle$              \\
		 & $D(n_c, n_d) \equiv 1 \pmod{2}$
		 & $|\Psi^+_L\rangle$              \\
		\midrule
		\multirow{2}{*}{$n_c + n_d \equiv N \pmod{2N}$}
		 & $D(n_c, n_d) \equiv 0 \pmod{2}$
		 & $|\Phi^-_L\rangle$              \\
		 & $D(n_c, n_d) \equiv 1 \pmod{2}$
		 & $|\Psi^-_L\rangle$              \\
		\bottomrule
	\end{tabular}
	\label{table:bsm_rs_cat_without_loss}
\end{table}

\noindent
Here,
\begin{equation}
	D(n_c, n_d) =
	\begin{cases}
		\operatorname{round}\left(\frac{N}{\pi}\arccos\frac{n_c - n_d}{n_c + n_d}\right)
		 & (n_c + n_d \neq 0) \\
		\operatorname{round}\left(N / 2\right)
		 & (n_c + n_d = 0)
	\end{cases},
\end{equation}
and $\operatorname{round}(\cdot)$ denotes rounding to the nearest integer. In the following, we show how this identification method distinguishes the Bell states and enables near-deterministic BSM for sufficiently large amplitudes $\alpha$.

\subsection{Discrimination between \texorpdfstring{$+$}{+} and \texorpdfstring{$-$}{-}}
\label{sec:bsm_rs_cat_1}

We show that the total photon number modulo $2N$ distinguishes the two sets $\{|\Phi^+_L\rangle, |\Psi^+_L\rangle\}$ and $\{|\Phi^-_L\rangle, |\Psi^-_L\rangle\}$. From the definitions in \cref{eq:fock_rs_cat}, we obtain
\begin{align}
	|0_L 0_L\rangle & = \sum_{k=0}^\infty \sum_{l=0}^\infty f_k f_l |kN, lN\rangle, \nonumber        \\
	|1_L 1_L\rangle & = \sum_{k=0}^\infty \sum_{l=0}^\infty (-1)^{k+l} f_k f_l |kN, lN\rangle,       \\
	|0_L 1_L\rangle & = \sum_{k=0}^\infty \sum_{l=0}^\infty (-1)^l f_k f_l |kN, lN\rangle, \nonumber \\
	|1_L 0_L\rangle & = \sum_{k=0}^\infty \sum_{l=0}^\infty (-1)^k f_k f_l |kN, lN\rangle. \nonumber
\end{align}
% \begin{equation}
% 	\begin{aligned}
% 		|0_L 0_L\rangle & = \sum_{k=0}^\infty \sum_{l=0}^\infty f_k f_l |kN, lN\rangle,            \\
% 		|1_L 1_L\rangle & = \sum_{k=0}^\infty \sum_{l=0}^\infty (-1)^{k+l} f_k f_l |kN, lN\rangle, \\
% 		|0_L 1_L\rangle & = \sum_{k=0}^\infty \sum_{l=0}^\infty (-1)^l f_k f_l |kN, lN\rangle,     \\
% 		|1_L 0_L\rangle & = \sum_{k=0}^\infty \sum_{l=0}^\infty (-1)^k f_k f_l |kN, lN\rangle.
% 	\end{aligned}
% \end{equation}
Substituting these expressions into \cref{eq:bell_state} yields
\begin{equation}
	\label{eq:bell_fock_basis}
	\begin{aligned}
		|\Phi^+_L\rangle & = \sqrt2 \sum_{k+l\equiv 0 \pmod{2}} f_k f_l |kN, lN\rangle,        \\
		|\Phi^-_L\rangle & = \sqrt2 \sum_{k+l\equiv 1 \pmod{2}} f_k f_l |kN, lN\rangle,        \\
		|\Psi^+_L\rangle & = \sqrt2 \sum_{k+l\equiv 0 \pmod{2}} (-1)^l f_k f_l |kN, lN\rangle, \\
		|\Psi^-_L\rangle & = \sqrt2 \sum_{k+l\equiv 1 \pmod{2}} (-1)^l f_k f_l |kN, lN\rangle.
	\end{aligned}
\end{equation}
\Cref{eq:bell_fock_basis} shows that $|\Phi^+_L\rangle$ and $|\Psi^+_L\rangle$ contain only terms where $k + l$ is even. The total photon number in the input modes is therefore always
\begin{equation}
	n_a + n_b = (k+l)N = 2mN.
\end{equation}
By contrast, $|\Phi^-_L\rangle$ and $|\Psi^-_L\rangle$ contain only odd $k + l$ terms, and the total photon number in the input modes is always
\begin{equation}
	n_a + n_b = (k+l)N = (2m+1)N.
\end{equation}
Since the total photon number is conserved by the HBS, the two sets $\{|\Phi^+_L\rangle, |\Psi^+_L\rangle\}$ and $\{|\Phi^-_L\rangle, |\Psi^-_L\rangle\}$ can be distinguished by checking whether the total number of detected photons satisfies $n_c + n_d \equiv 0 \pmod{2N}$ or $n_c + n_d \equiv N \pmod{2N}$.

\subsection{Discrimination between \texorpdfstring{$\Phi$}{Phi} and \texorpdfstring{$\Psi$}{Psi}}
\label{sec:bsm_rs_cat_2}

We next show that the phase information associated with the rotational symmetry of RS-cat codes is extracted from the detected photon-number pattern, thereby enabling the discrimination of the two sets $\{|\Phi^\pm_L\rangle\}$ and $\{|\Psi^\pm_L\rangle\}$. From the definition in \cref{eq:def_rs_cat}, we obtain
\begin{equation}
	\begin{aligned}
		|\Phi^\pm_L\rangle
		\propto \sum_{m=0}^{N-1} \sum_{m'=0}^{N-1}(|
		 & e^{\frac{2m\pi i}{N}}\alpha\rangle |e^{\frac{2m'\pi i}{N}}\alpha\rangle                \\
		 & \pm |e^{\frac{(2m+1)\pi i}{N}}\alpha\rangle |e^{\frac{(2m'+1)\pi i}{N}}\alpha\rangle), \\
		|\Psi^\pm_L\rangle
		\propto \sum_{m=0}^{N-1} \sum_{m'=0}^{N-1}(|
		 & e^{\frac{2m\pi i}{N}}\alpha\rangle |e^{\frac{(2m'+1)\pi i}{N}}\alpha\rangle            \\
		 & \pm |e^{\frac{(2m+1)\pi i}{N}}\alpha\rangle |e^{\frac{2m'\pi i}{N}}\alpha\rangle).
	\end{aligned}
\end{equation}
These expressions can be rewritten as
\begin{equation}
	\begin{aligned}
		|\Phi^+_L\rangle & \propto \sum_{m=0}^{2N-1} \sum_{q=0}^{N-1}
		|e^{\frac{m\pi i}{N}}\alpha\rangle
		|e^{\frac{(m+2q)\pi i}{N}}\alpha\rangle,                      \\
		|\Phi^-_L\rangle & \propto \sum_{m=0}^{2N-1} \sum_{q=0}^{N-1}
		(-1)^m
		|e^{\frac{m\pi i}{N}}\alpha\rangle
		|e^{\frac{(m+2q)\pi i}{N}}\alpha\rangle,                      \\
		|\Psi^+_L\rangle & \propto \sum_{m=0}^{2N-1} \sum_{q=0}^{N-1}
		|e^{\frac{m\pi i}{N}}\alpha\rangle
		|e^{\frac{(m+2q+1)\pi i}{N}}\alpha\rangle,                    \\
		|\Psi^-_L\rangle & \propto \sum_{m=0}^{2N-1} \sum_{q=0}^{N-1}
		(-1)^m
		|e^{\frac{m\pi i}{N}}\alpha\rangle
		|e^{\frac{(m+2q+1)\pi i}{N}}\alpha\rangle.
	\end{aligned}
\end{equation}
Thus, the coherent-state expansions show that $|\Phi^\pm_L\rangle$ and $|\Psi^\pm_L\rangle$ are characterized by distinct sets of relative phases, namely $S_{\mathrm{even}} = \bigl\{\tfrac{2q\pi}{N}\bigr\}_{q=0}^{N-1}$ and $S_{\mathrm{odd}} = \bigl\{\tfrac{(2q+1)\pi}{N}\bigr\}_{q=0}^{N-1}$, respectively.

We now show how this phase information is reflected in the output photon-number pattern. From \cref{eq:bs_coherent}, we have
\begin{equation}
	\begin{aligned}
		\label{eq:after_hbs}
		U_{\mathrm{HBS}}|e^{i\theta}\alpha\rangle |e^{i(\theta + \delta)}\alpha\rangle
		= \left|\tfrac{1 + e^{i\delta}}{\sqrt2}e^{i\theta}\alpha\right\rangle
		\left|\tfrac{1 - e^{i\delta}}{\sqrt2}e^{i\theta}\alpha\right\rangle.
	\end{aligned}
\end{equation}
Thus, the detected photon numbers satisfy
\begingroup
\small
\begin{equation}
	\begin{aligned}
		\label{eq:nc_nd_poisson}
		n_c & \sim \mathrm{Poisson}(\mu_c), \quad \mu_c = \left| \frac{1 + e^{i\delta}}{\sqrt{2}} e^{i\theta}\alpha \right|^2 = (1 + \cos\delta)\alpha^2, \\
		n_d & \sim \mathrm{Poisson}(\mu_d), \quad \mu_d = \left| \frac{1 - e^{i\delta}}{\sqrt{2}} e^{i\theta}\alpha \right|^2 = (1 - \cos\delta)\alpha^2.
	\end{aligned}
\end{equation}
\endgroup
For two such independent Poisson distributions, the conditional distribution of $n_c$ given the total photon number $n_{c+d} = n_c + n_d$ is
\begin{equation}
	n_c \mid n_{c+d} \sim \mathrm{Binomial}(n_{c+d}, \; p),
\end{equation}
where $p = \frac{\mu_c}{\mu_c + \mu_d} = \frac{1 + \cos\delta}{2}$. The expectation and variance are given by
\begin{equation}
	\begin{aligned}
		E[n_c | n_{c+d}]
		 & = n_{c+d} \cdot p =  \frac{n_{c+d} \; (1+ \cos\delta)}{2},     \\
		\operatorname{Var}(n_c | n_{c+d})
		 & = n_{c+d} \cdot p\;(1-p) = \frac{n_{c+d}\cdot\sin^2\delta}{4}.
	\end{aligned}
\end{equation}
We now define the normalized photon-number difference
\begin{equation}
	X =
	\begin{cases}
		\frac{2}{n_{c+d}} n_c - 1 = \frac{n_c - n_d}{n_c + n_d} & (n_{c+d} \neq 0) \\
		0                                                       & (n_{c+d} = 0)
	\end{cases}.
\end{equation}
For $n_{c+d} \neq 0$, this quantity provides an estimator of the relative phase through $\cos\delta$. Its conditional mean and variance are
\begingroup
\small
\begin{equation}
	\begin{aligned}
		E[X|n_{c+d}]
		 & = \begin{cases}
			     \frac{2}{n_{c+d}}E[n_c|n_{c+d}] - 1
			     = \cos\delta
			      & (n_{c+d} \neq 0) \\
			     0
			      & (n_{c+d} = 0)
		     \end{cases},                   \\
		\operatorname{Var}(X | n_{c+d})
		 & = \begin{cases}
			     \frac{4}{n_{c+d}^2} \operatorname{Var}(n_c | n_{c+d})
			     =  \frac{\sin^2\delta}{n_{c+d}}
			      & (n_{c+d} \neq 0) \\
			     0
			      & (n_{c+d} = 0)
		     \end{cases}.
	\end{aligned}
\end{equation}
\endgroup

\begin{figure}[htbp]
	\centering
	\includegraphics[width=0.9\linewidth]{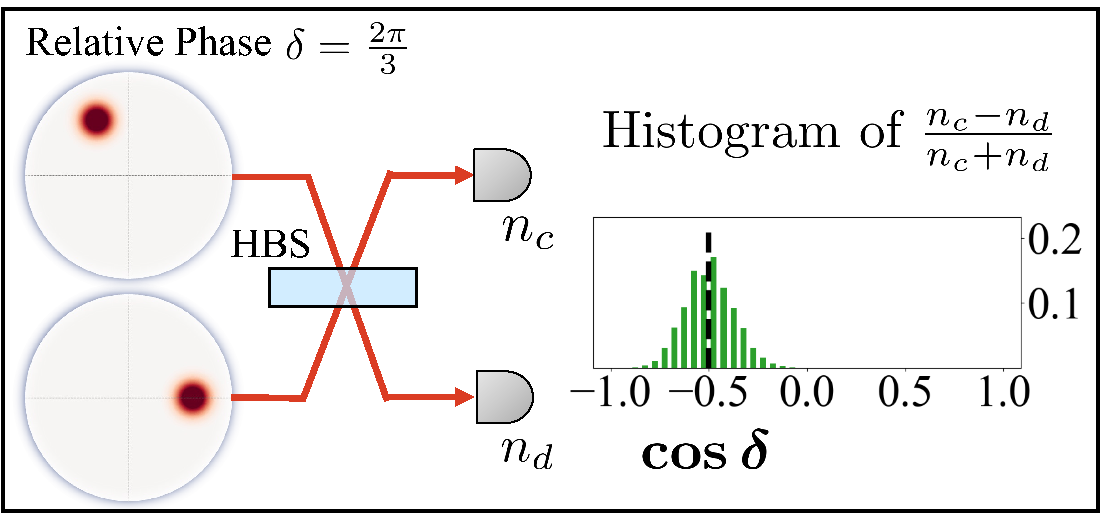}
	\caption{Histogram of $X$ for coherent states $|\alpha\rangle |e^{2\pi i/3}\alpha\rangle$ with relative phase $\delta=\tfrac{2\pi}{3}$. The histogram is concentrated around the value indicated by the vertical dashed line, $\cos\delta=-\frac{1}{2}$.}
	\label{fig:histogram_rho}
\end{figure}

\begin{figure*}[tb]
	\includegraphics[width=\linewidth]{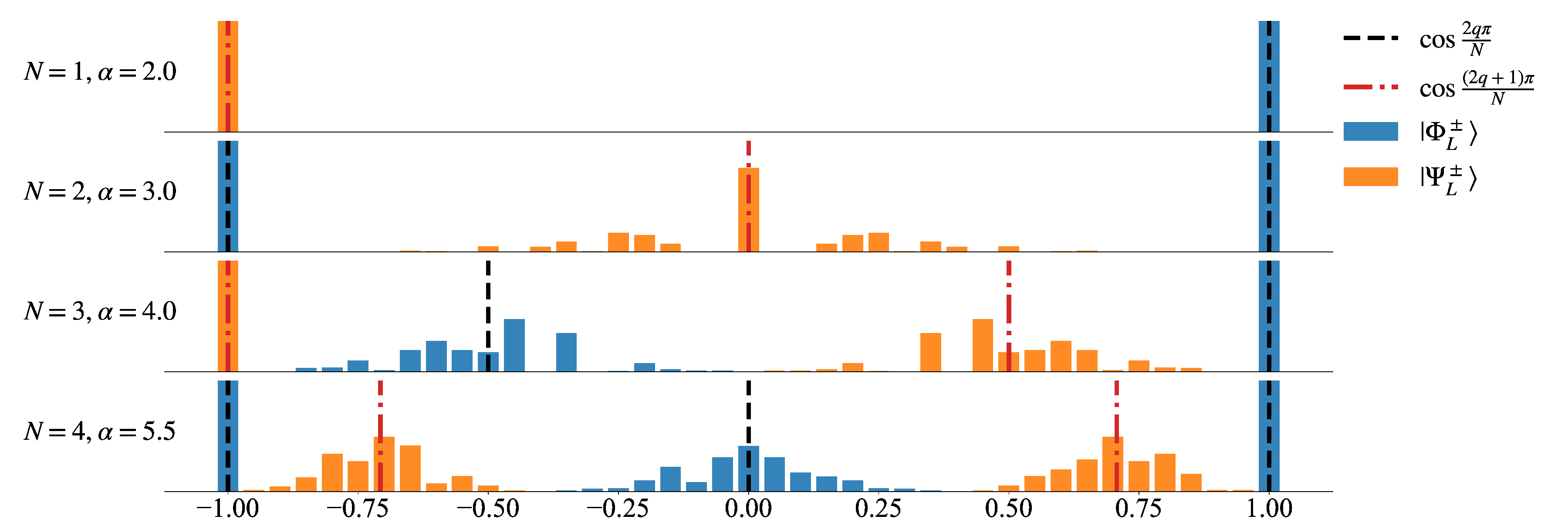}
	\caption{Histograms of $X$ for the Bell basis states of the RS-cat code. The blue and orange bars correspond to $|\Phi_L^\pm\rangle$ and $|\Psi_L^\pm\rangle$, respectively. The histograms are concentrated around the values indicated by the vertical dashed lines, with the black lines denoting $\{\cos\tfrac{2q\pi}{N}\}_{q=0}^{N-1}$ and the red lines denoting $\{\cos\tfrac{(2q+1)\pi}{N}\}_{q=0}^{N-1}$.}
	\label{fig:histogram_cat}
\end{figure*}

\noindent
The marginal distribution of $X$ can be written as \vspace{-1em}
\begin{equation}
	P(X) = \sum_{n_{c+d}=0}^\infty P(n_{c+d}) \cdot P(X|n_{c+d}),
\end{equation}
and \vspace{-1em}
\begingroup
\small
\begin{align}
	E[X]
	 & = \sum_{n_{c+d}=0}^\infty P(n_{c+d})\cdot E[X|n_{c+d}] \nonumber                  \\
	 & = 0 + \sum_{n_{c+d}=1}^\infty P(n_{c+d})\cdot \cos\delta \nonumber                \\
	 & = \left(1 - e^{-2\alpha^2}\right) \cos\delta \nonumber                            \\
	 & \xrightarrow{\alpha \to \infty} \; \cos\delta,                                    \\
	\operatorname{Var}(X)
	 & = E[\operatorname{Var}(X|n_{c+d})]
	+ \operatorname{Var}(E[X|n_{c+d}]) \nonumber                                         \\
	 & = E[\operatorname{Var}(X|n_{c+d})] \nonumber                                      \\
	 & \quad + E\left[(E[X|n_{c+d}])^2\right]
	- \left(E[E[X|n_{c+d}]])\right)^2 \nonumber                                          \\
	 & = \sum_{n_{c+d}=1}^\infty \!\! P(n_{c+d})
	\cdot \frac{\sin^2\delta}{n_{c+d}} \nonumber                                         \\
	 & \quad + \sum_{n_{c+d}=1}^\infty \!\! P(n_{c+d}) \cos^2\delta
	- \left(\sum_{n_{c+d}=1}^\infty \!\! P(n_{c+d})\right)^2 \!\! \cos^2\delta \nonumber \\
	 & \simeq \left(1 - e^{-2\alpha^2}\right)\frac{\sin^2\delta}{2\alpha^2}
	+ \left(1 - e^{-2\alpha^2}\right) \cos^2\delta \nonumber                             \\
	 & \quad - \left(1 - e^{-2\alpha^2}\right)^2 \cos^2\delta
	\qquad\qquad\qquad(\alpha \gg  1) \nonumber                                          \\
	 & \xrightarrow{\alpha \to \infty} \; 0. \nonumber
\end{align}
\endgroup
Thus, the distribution of $X$ concentrates around $\cos\delta$. For example, \cref{fig:histogram_rho} shows the distribution of $X$ for the input state $\left|\alpha\right\rangle \left|e^{2\pi i/3}\alpha\right\rangle$.
As $\alpha$ becomes sufficiently large, the distribution develops an increasingly sharp peak around $\cos\delta$.

As shown in \cref{appendix:additional_proof}, when the amplitude $\alpha$ is sufficiently large, the distribution of $X$ for each Bell basis state of the RS-cat code can be approximated by a sum of the corresponding coherent-state distributions. It therefore concentrates around $\{\cos\tfrac{2q\pi}{N}\}_{q=0}^{N-1}$ for $|\Phi^\pm_L\rangle$ and $\{\cos\tfrac{(2q+1)\pi}{N}\}_{q=0}^{N-1}$ for $|\Psi^\pm_L\rangle$, respectively. This behavior is confirmed numerically in \cref{fig:histogram_cat}. We therefore define
\begin{equation}
	D(n_c, n_d) = \operatorname{round}\left(\frac{N}{\pi}\arccos X\right),
\end{equation}
whose parity distinguishes between $\{|\Phi^\pm_L\rangle\}$ and $\{|\Psi^\pm_L\rangle\}$. The small-amplitude behavior is discussed numerically in \cref{appendix:histogram_small_alpha}.

\subsection{Performance of our BSM protocol}
\label{sec:without_loss}

\begin{table*}[htbp]
	\centering
	\caption{Decision rules for Bell-state identification with RS-cat codes considering photon loss.}
	\begin{tabular}{ccc}
		\toprule
		\multicolumn{2}{c}{Measurement Pattern}
		 & \quad Decision   \quad                       \\
		\cmidrule(r){1-2} \cmidrule(l){3-3}
		\multirow{2}{*}{\quad $n_c + n_d \pmod{2N} \in \{0, N+1, \dots, 2N-1\}$ \quad}
		 & \quad $D(n_c, n_d) \equiv 0 \pmod{2} \quad $
		 & $|\Phi^+_L\rangle$                           \\
		 & $D(n_c, n_d) \equiv 1 \pmod{2}$
		 & $|\Psi^+_L\rangle$                           \\
		\midrule
		\multirow{2}{*}{$n_c + n_d \pmod{2N} \in \{1, \dots N\}$}
		 & \quad $D(n_c, n_d) \equiv 0 \pmod{2} \quad $
		 & $|\Phi^-_L\rangle$                           \\
		 & \quad $D(n_c, n_d) \equiv 1 \pmod{2} \quad $
		 & $|\Psi^-_L\rangle$                           \\
		\bottomrule
	\end{tabular}
	\label{table:bsm_rs_cat_with_loss}
\end{table*}

We numerically evaluate the performance of our BSM protocol for RS-cat codes, first considering the lossless case. In this work, we focus on RS-cat codes with symmetry orders $N=1,2,3,$ and $4$, although the proposed protocol is applicable to arbitrary $N$.

\begin{figure}[htbp]
	\centering
	\begin{tikzpicture}
		\node (img) {\includegraphics[width=0.93\linewidth]{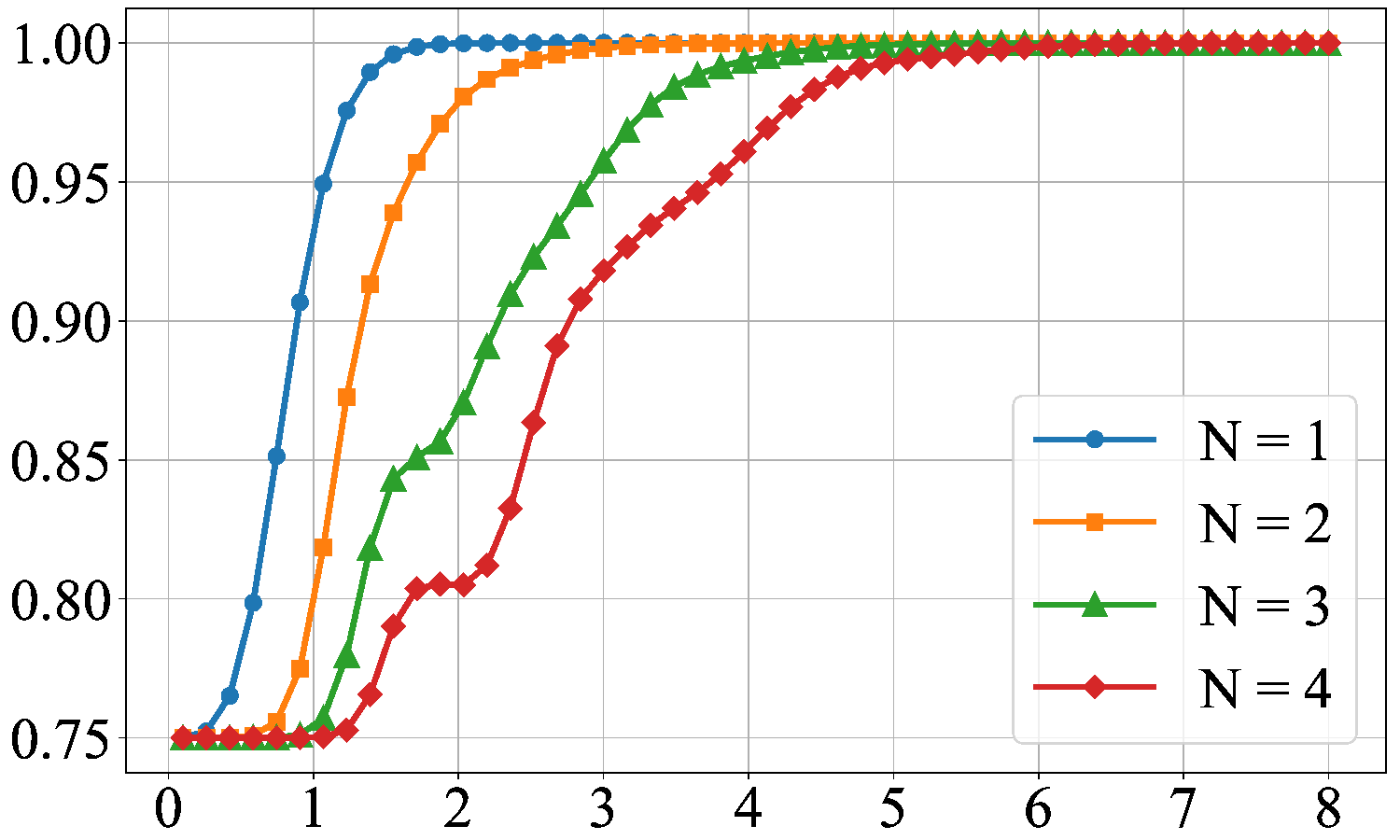}};
		\node[above=-1mm of img, xshift=1em] {Without Photon Loss};
		\node[below=-2mm of img, xshift=1em] {Amplitude $\alpha$};
		\node[left=1mm of img.west, rotate=90, anchor=center, xshift=1em] {Success Probability};
	\end{tikzpicture}
	\caption{Success probability of the BSM method for RS-cat codes under lossless conditions. For each symmetry order $N$, the BSM becomes near-deterministic as the amplitude $\alpha$ increases.}
	\label{fig:success_rate}
\end{figure}
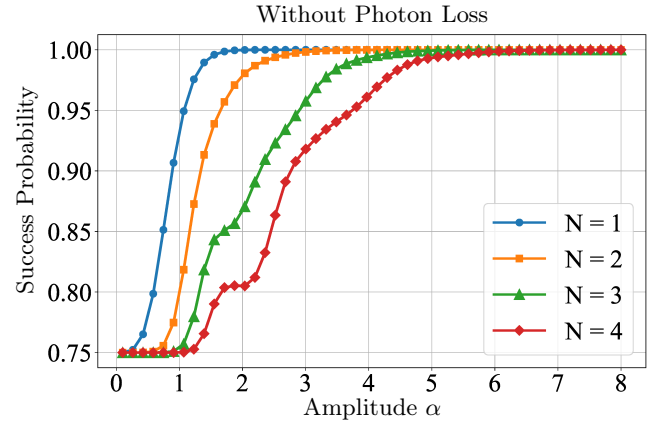

\Cref{fig:success_rate} shows the success probability of the BSM as a function of $\alpha$. The blue, orange, green, and red curves correspond to RS-cat codes with $N = 1, 2, 3,$ and $4$, respectively. These results indicate that the proposed BSM protocol can become near-deterministic for any symmetry order $N$, provided that the amplitude $\alpha$ is sufficiently large. We also observe that larger symmetry orders require larger amplitudes to reach the near-deterministic regime.

\begin{figure}[htbp]
	\centering
	\begin{tikzpicture}
		\node (img) {\includegraphics[width=0.93\linewidth]{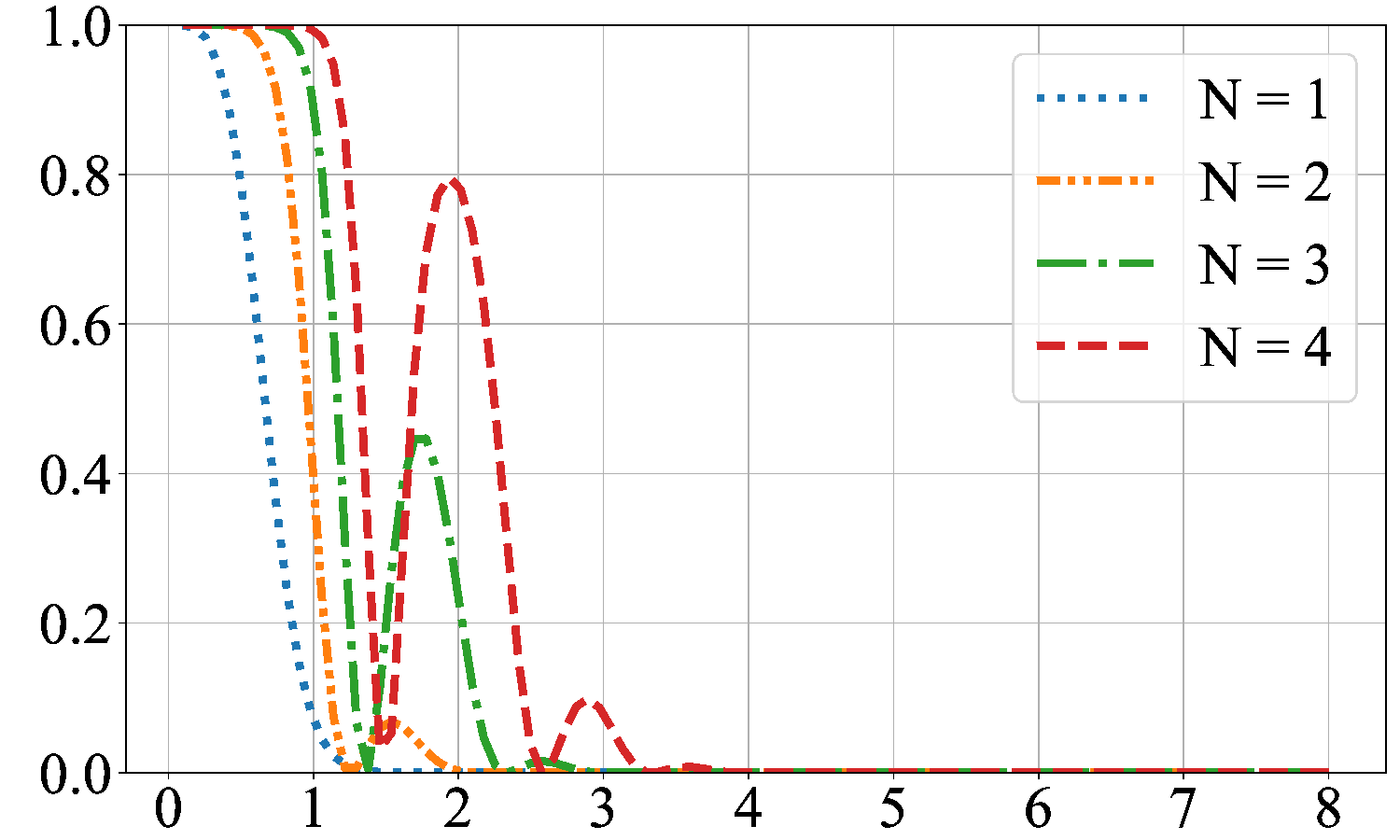}};
		\node[below=-2mm of img, xshift=1em] {Amplitude $\alpha$};
		\node[left=1mm of img.west, rotate=90, anchor=center, xshift=1em] {$|\langle \Phi^+_L | \Psi^+_L\rangle|^2$};
	\end{tikzpicture}
	\caption{Orthogonality of the Bell basis of RS-cat codes.}
	\label{fig:orthogonality}
\end{figure}

As is particularly evident for $N=3$ and $N=4$ around $\alpha=2$, the growth of the success probability temporarily slows down. We attribute this behavior to the imperfect orthogonality of the Bell basis states of the RS-cat codes, which reduces their distinguishability and consequently limits the BSM success probability. \Cref{fig:orthogonality} shows the orthogonality of the Bell basis of RS-cat codes as a function of the amplitude $\alpha$. The overlap between the different Bell states actually does not decrease monotonically with $\alpha$, but instead exhibits revivals at certain values of $\alpha$.

The horizontal and vertical axes represent the amplitude $\alpha$ and $|\langle \Phi^+_L | \Psi^+_L \rangle|^2$, respectively. We plot this particular overlap because, among the Bell basis states of the RS-cat codes, only $|\Phi^+_L\rangle$ and $|\Psi^+_L\rangle$ are non-orthogonal, whereas all other pairs are orthogonal. This temporary increase in the overlap is consistent with the slowdown in the growth of the BSM success probability for $N=3$ and $N=4$ around $\alpha = 2$.

\subsection{Performance of BSM with photon loss}
\label{sec:with_loss}

We now turn to the effect of photon loss. The decision rules for discriminating between $\{|\Phi^\pm_L\rangle\}$ and $\{|\Psi^\pm_L\rangle\}$ in the loss-free case remain applicable in the presence of photon loss. This is because the normalized photon-number difference $X$ is relatively insensitive to photon loss for sufficiently large $\alpha$, allowing the parity of $D(n_c,n_d)$ to remain a robust indicator of the relative-phase class. On the other hand, as with the RS-cat code itself, the discrimination between $\{|\Phi^+_L\rangle,|\Psi^+_L\rangle\}$ and $\{|\Phi^-_L\rangle,|\Psi^-_L\rangle\}$ is robust against up to $N-1$ photon-loss events. The corresponding decision rules are therefore extended as summarized in \cref{table:bsm_rs_cat_with_loss}.

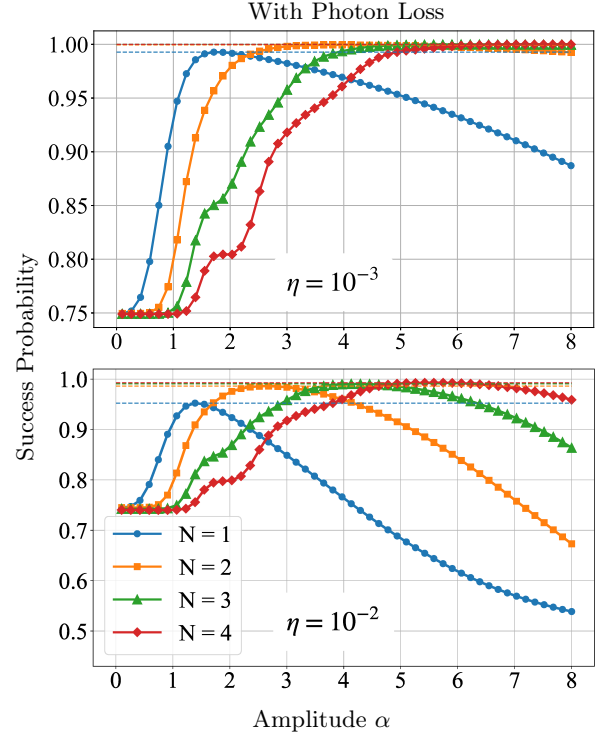
\begin{figure}[htbp]
	\centering
	\begin{tikzpicture}
		\node (img1) {\includegraphics[width=0.85\linewidth]{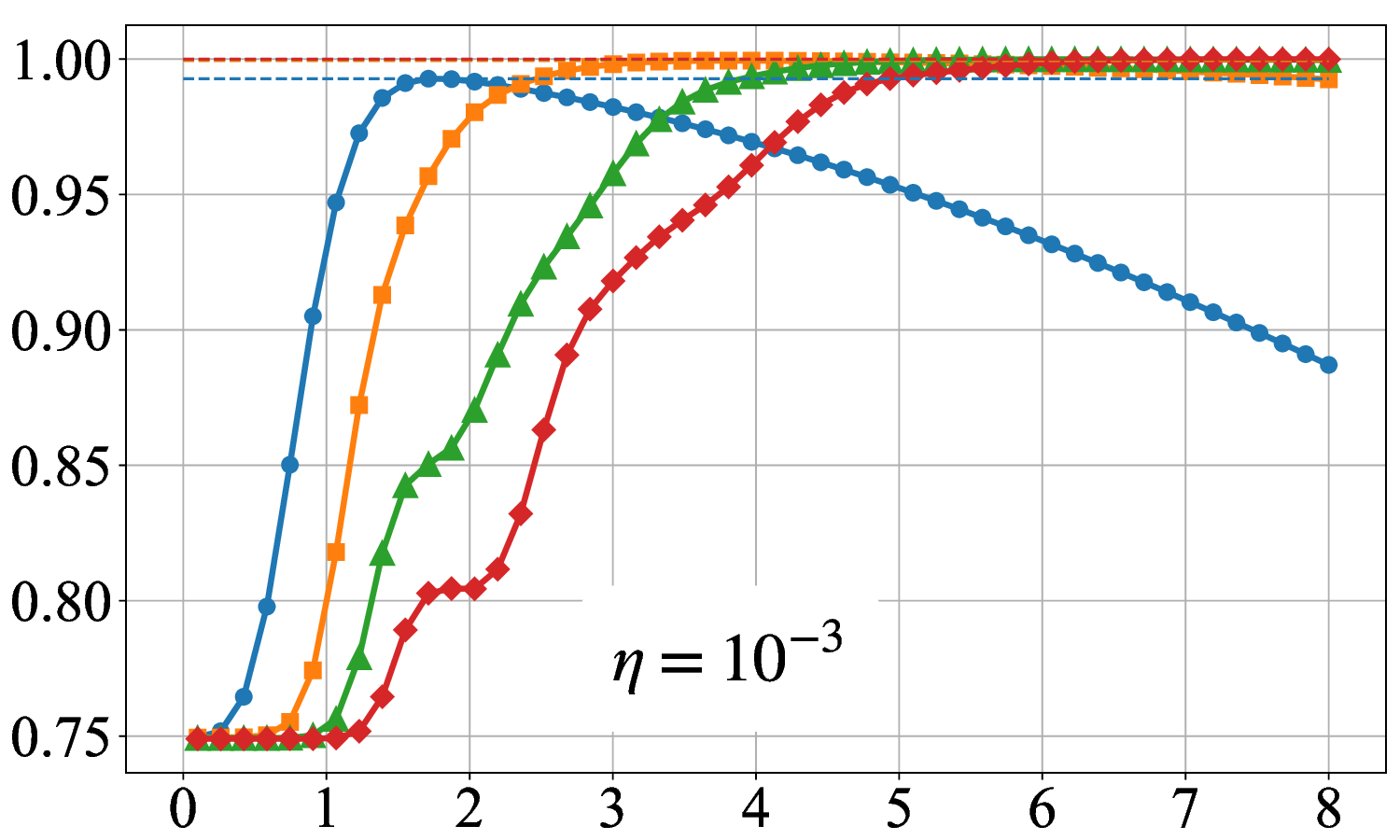}};
		\node[above=-2mm of img1, xshift=1em] {With Photon Loss};
		\node (img2) [below=-0.5em of img1] {\includegraphics[width=0.85\linewidth]{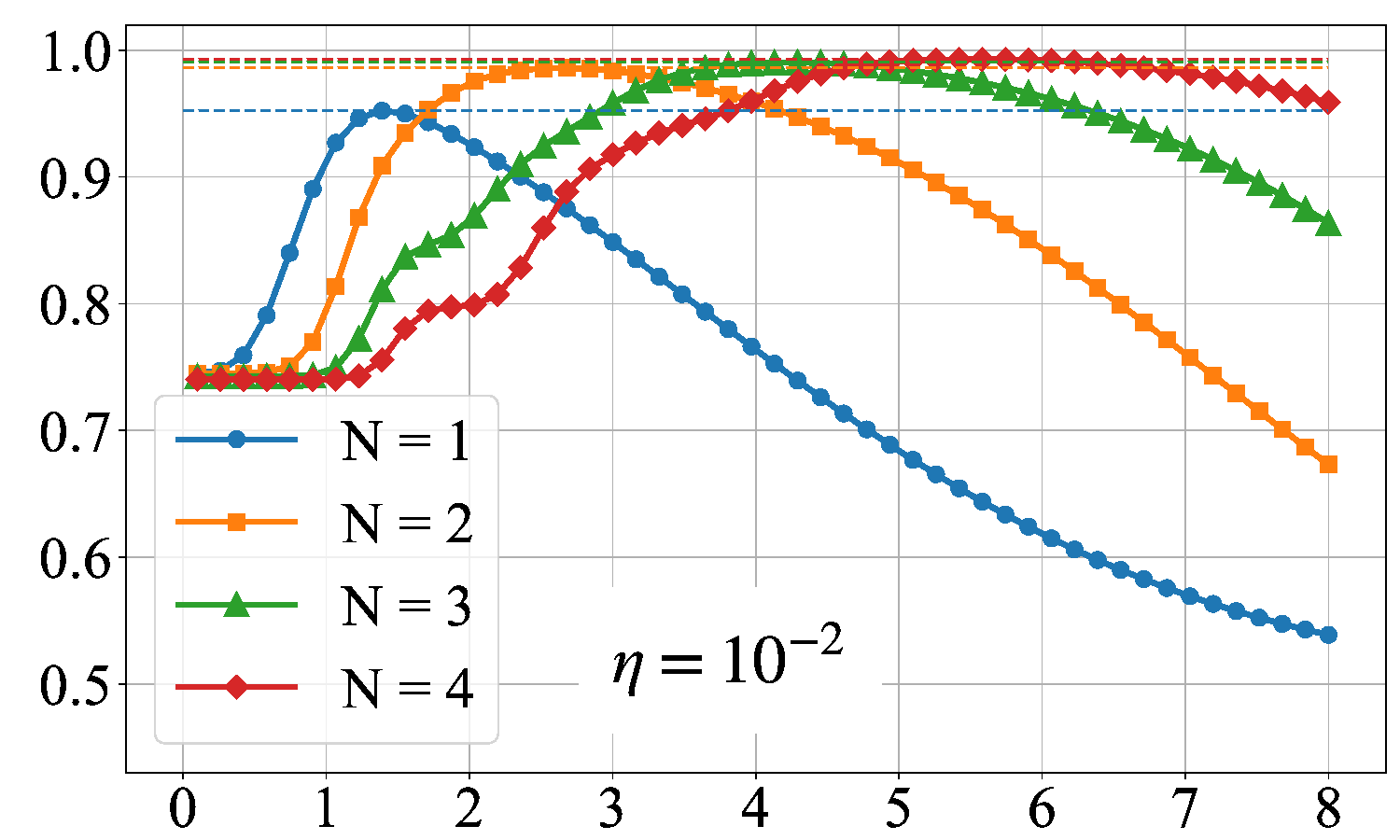}};
		\node [below=0em of img2] {Amplitude $\alpha$};
		\node[left=-1.5mm of img2.west, rotate=90, anchor=south] at ($(img1.north west)!0.5!(img2.south west)$) {Success Probability};
	\end{tikzpicture}
	\caption{Success probability of the BSM under photon loss. The upper and lower panels show the results for photon loss rates $\eta = 10^{-3}$ and $\eta = 10^{-2}$, respectively.}
	\label{fig:success_rate_with_loss}
\end{figure}
\begin{figure}[htbp]
	\centering
	\begin{tikzpicture}
		\node (img) {\includegraphics[width=0.85\linewidth]{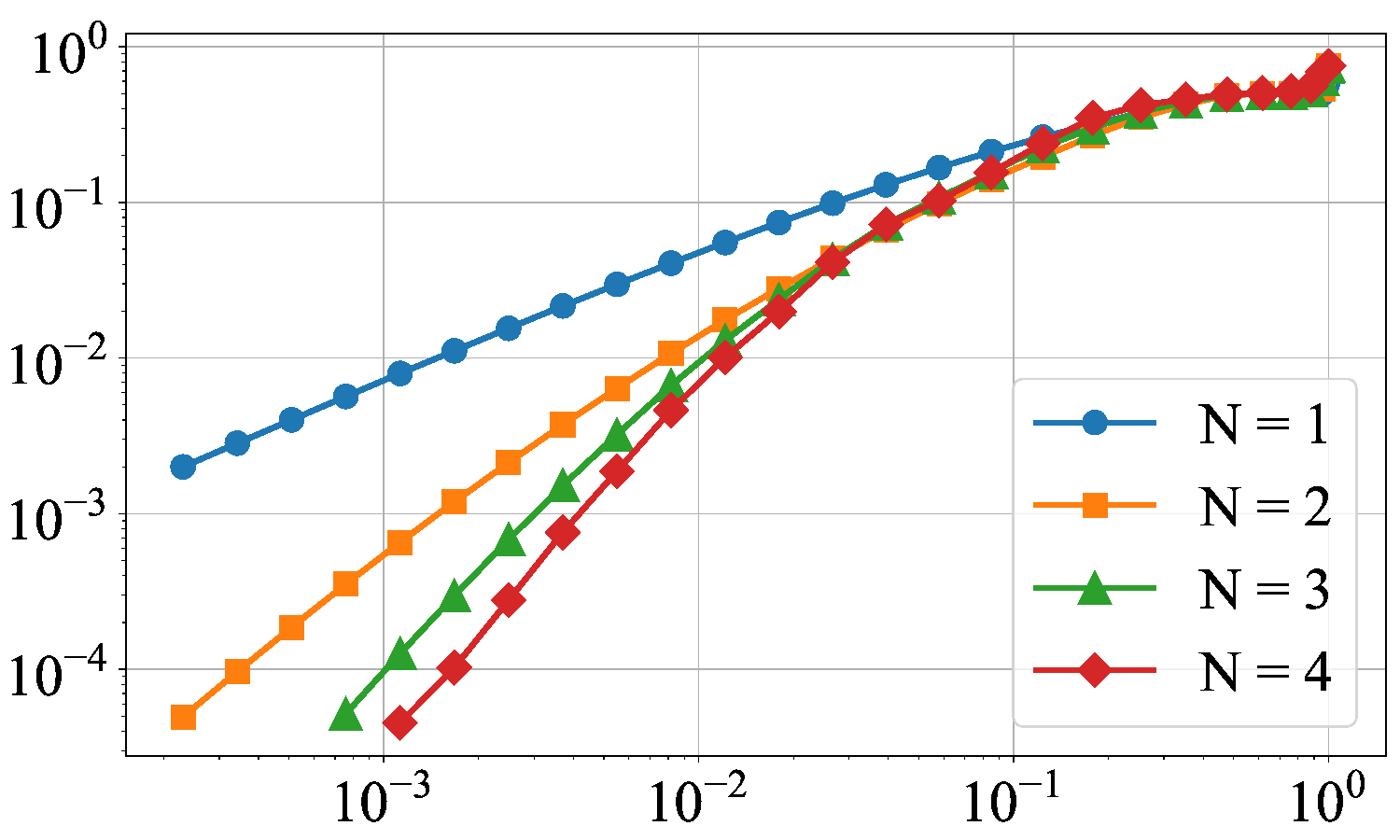}};
		\node[below=-2mm of img, xshift=1em] {Photon loss rate $\eta$};
		\node[left=1mm of img.west, rotate=90, anchor=center, xshift=1em] {Failure Rate};
	\end{tikzpicture}
	\caption{Failure rate of the BSM for RS-cat codes as a function of the photon loss rate $\eta$, where the amplitude $\alpha$ is optimized for each pair of $N$ and $\eta$. The curves show the results for symmetry orders $N = 1, 2, 3,$ and $4$. For $\eta \lesssim 0.5$, increasing the symmetry order $N$ reduces the failure rate.}
	\label{fig:success_rate_loss}
\end{figure}

We evaluate the robustness of our BSM protocol against photon loss by numerical simulation. \Cref{fig:success_rate_with_loss} shows the success probabilities for photon loss rates $\eta = 10^{-3}$ and $\eta = 10^{-2}$. Since the mean photon number scales as $|\alpha|^2$, larger amplitudes make the RS-cat states more vulnerable to photon loss, while improving the distinguishability of the Bell states. Thus, in the presence of photon loss, the optimal amplitude $\alpha$ is determined by both the loss rate $\eta$ and the symmetry order $N$.

We also examine the dependence of the BSM failure rate on the photon loss rate $\eta$, as shown in \cref{fig:success_rate_loss}. For each pair of $N$ and $\eta$, we optimize the amplitude $\alpha$ and plot the resulting minimum failure rate. The results indicate that, for $\eta \lesssim 0.5$, the failure rate decreases as the symmetry order $N$ increases. Thus, higher-order RS-cat codes provide an advantage for BSM when the photon-loss rate is below a certain threshold.

\subsection{Improving the performance through post-selection}
\label{sec:post-selection}

\begin{figure}[htbp]
	\centering
	\includegraphics[width=\linewidth]{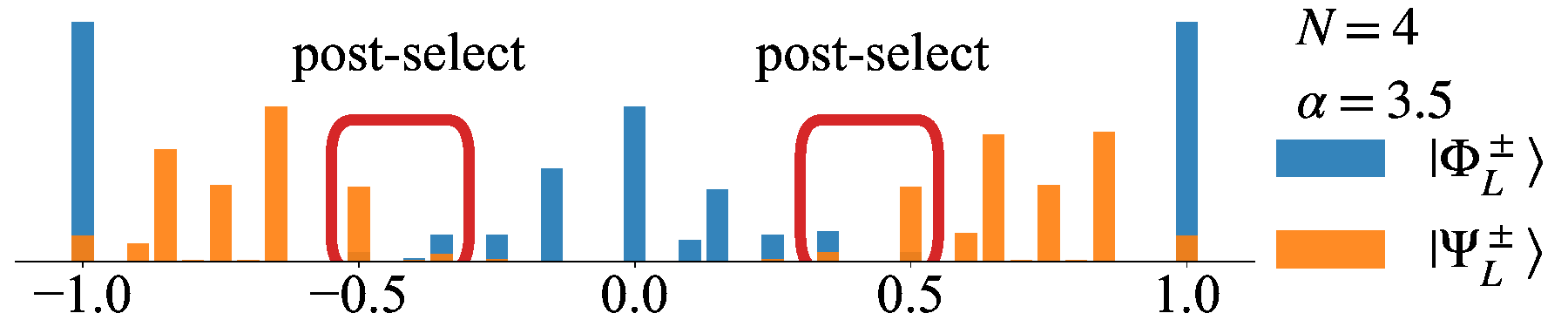}
	\caption{Histogram of $X$ for the $N=4$ RS-cat code for $\alpha=3.5$. The blue and orange bars represent $|\Phi_L^\pm\rangle$ and $|\Psi_L^\pm\rangle$, respectively. The regions marked by the red lines are rejected in the post-selection procedure, since they are close to the decision boundaries and lead to ambiguous Bell-state discrimination.}
	\label{fig:histogram_post_selection}
\end{figure}

Here we show the improvement of the success probability by introducing post-selection. In our protocol, one of the major causes of the failures is the overlap between the distributions of $X$ for different Bell states. These overlaps occur near the decision boundaries of the rounding operation in the discriminant $D(n_c, n_d)$, where the Bell-state assignment is ambiguous (see \cref{fig:histogram_post_selection}). We therefore introduce a threshold $\tau \in [0, 0.5]$ and reject the outcome when $n_c+n_d \neq 0$ and the following condition is satisfied:
\begin{equation}
	\left|D(n_c, n_d) -  \frac{N}{\pi}\arccos\frac{n_c - n_d}{n_c + n_d}\right| > 0.5 - \tau.
\end{equation}

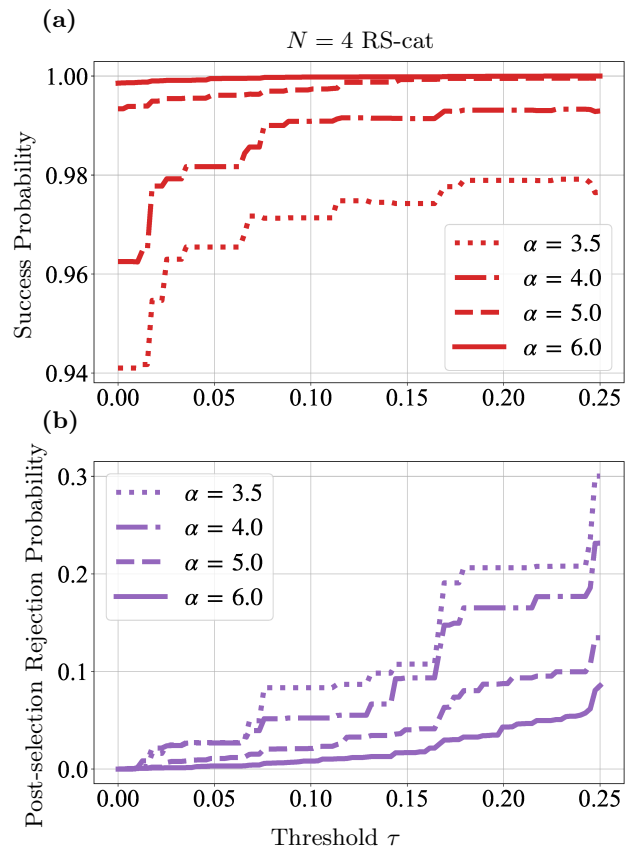
\begin{figure}[htbp]
	\centering
	\begin{tikzpicture}
		\node (img1) {\includegraphics[width=0.9\linewidth]{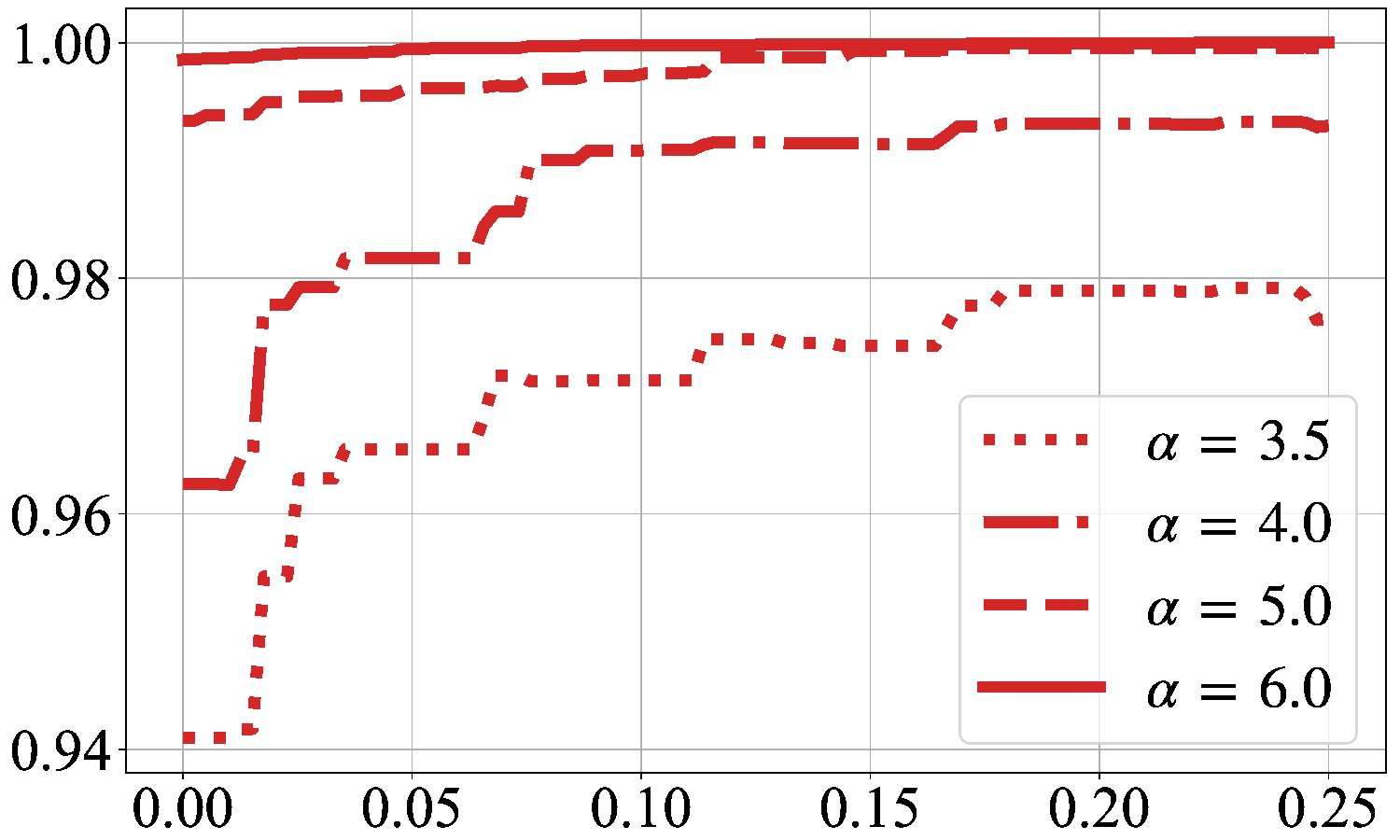}};
		\node[above=-1mm of img1, xshift=1em] {$N = 4$ RS-cat};
		\node[left=-1.5mm of img1, rotate=90, anchor=south] {Success Probability};
		\node[anchor=south west, xshift=0mm, yshift=1mm] at (img1.north west) {\textbf{(a)}};

		\node[below=4mm of img1] (img2) {\includegraphics[width=0.9\linewidth]{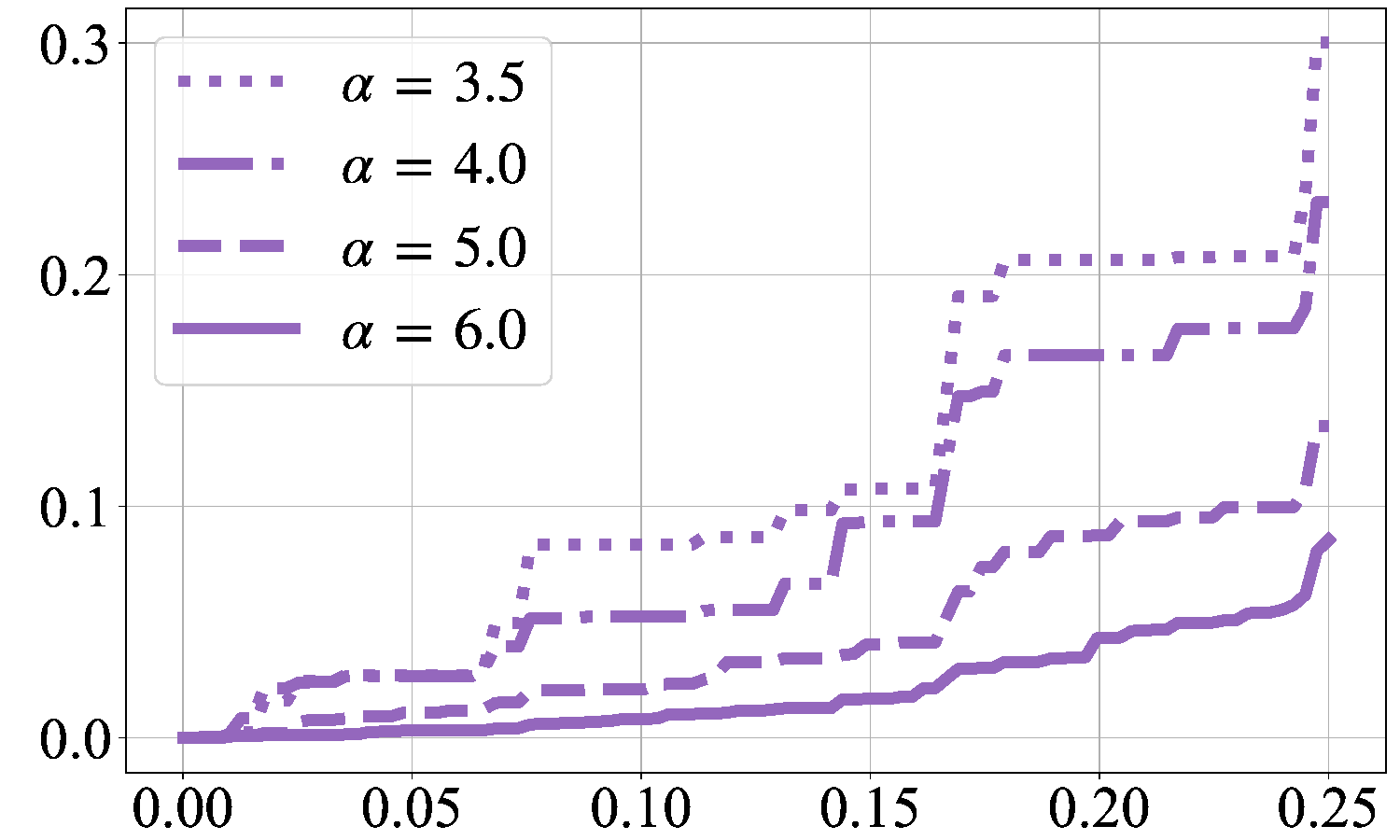}};
		\node[left=-3mm of img2, rotate=90, anchor=south] {Post-selection Rejection Probability};
		\node[below=0em of img2] {Threshold $\tau$};
		\node[anchor=south west, xshift=0mm, yshift=1mm] at (img2.north west) {\textbf{(b)}};
	\end{tikzpicture}
	\caption{Post-selection performance for the $N=4$ RS-cat code under lossless conditions, shown as functions of the threshold $\tau$ for amplitudes $\alpha = 3.5, 4.0, 5.0,$ and $6.0$. (a) BSM success probability. (b) Post-selection rejection probability, which quantifies the additional resource overhead induced by post-selection.}
	\label{fig:post_selection}
\end{figure}

\Cref{fig:post_selection} shows the BSM success probability and the post-selection rejection probability as functions of the threshold $\tau$. The results are shown for the $N=4$ RS-cat code with $\alpha = 3.5, 4.0, 5.0,$ and $6.0$. These results demonstrate that post-selection enhances the success probability at the expense of a higher rejection probability, i.e., increased resource overhead.
\section{Conclusion and Discussion}
\label{sec:conclusion}

In this work, we considered a BSM protocol for RS-cat codes that can be implemented using only a HBS and two PNRDs. The protocol identifies Bell states by extracting both the photon-number modulo information and the phase information inherent in RS-cat codes. Under ideal loss-free conditions, our numerical results show that the BSM succeeds deterministically for any symmetry order $N$, provided that the amplitude $\alpha$ is sufficiently large. We also investigated the effect of photon loss and found that, for loss rates $\eta \lesssim 0.5$, the failure probability decreases as the symmetry order $N$ increases. Furthermore, we showed that post-selection can improve the BSM performance.

\begin{figure}[tbp]
	\centering
	\begin{tikzpicture}
		\node (img) {\includegraphics[width=0.96\linewidth]{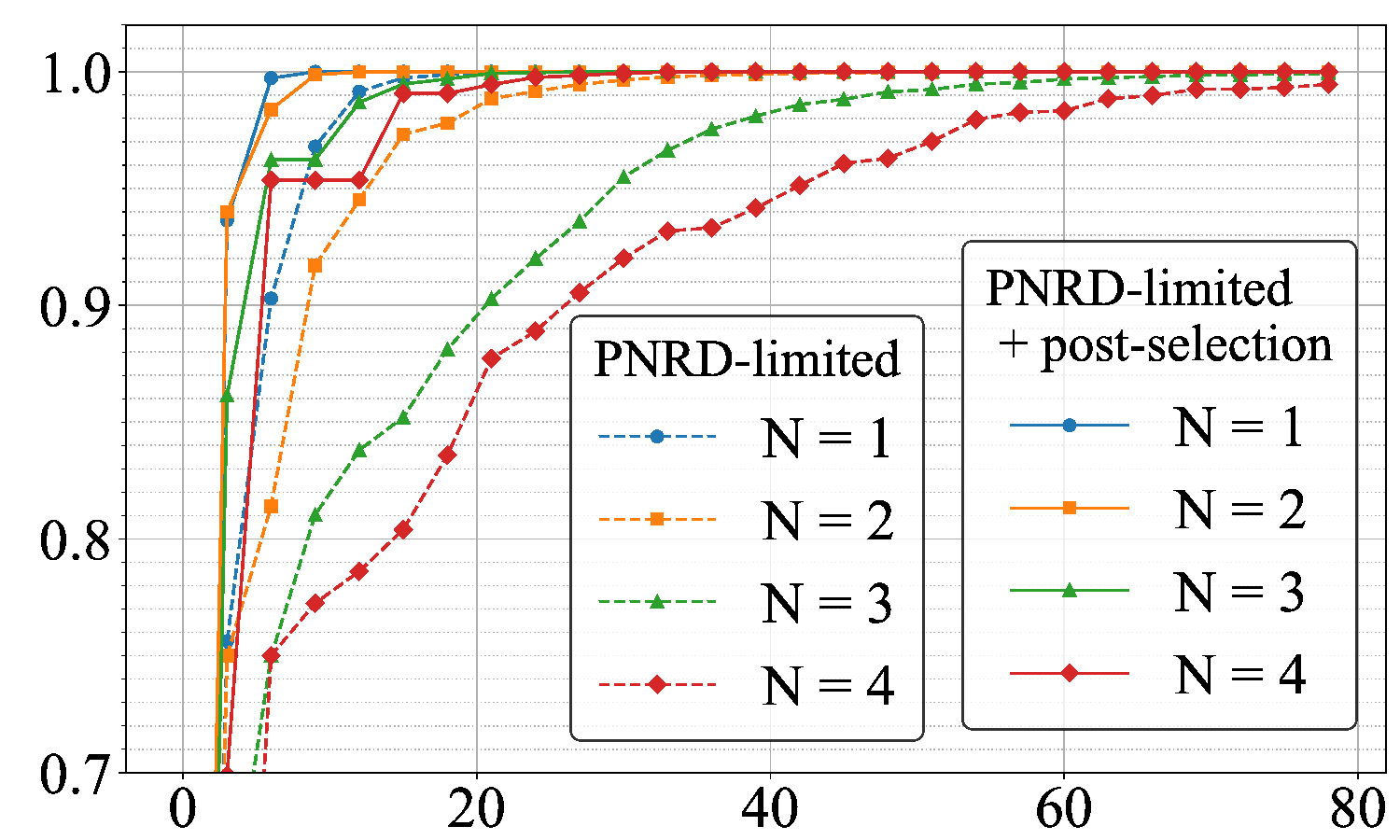}};
		\node[below=-1mm of img, xshift=1em] {PNRD resolution limit $n_{\rm PNRD}$};
		\node[left=-1.5mm of img.west, rotate=90, anchor=center, xshift=1em] {Success Probability};
	\end{tikzpicture}
	\caption{Maximum success probability of the BSM for RS-cat codes under finite PNRD resolution. For each resolution cutoff $n_{\rm PNRD}$, the success probability is maximized over the amplitude $\alpha$. The blue, orange, green, and red curves correspond to $N=1,2,3$, and $4$, respectively. Higher symmetry orders require larger $n_{\rm PNRD}$ to achieve near-deterministic BSM. Without post-selection, we regard any event in which one of the output photon numbers exceeds $n_{\rm PNRD}$ as a failure.}
	\label{fig:restrict_PNRD}
\end{figure}

While these results demonstrate the potential of our BSM protocol, its practical implementation is constrained by the finite photon-number-resolving capability of PNRDs. In fact, when one of the output photon numbers exceeds the PNRD resolution limit, the reliability of Bell-state discrimination is expected to degrade significantly. Based on this observation, we plot the maximum BSM success probability, optimized over $\alpha$, as a function of the PNRD resolution limit $n_{\rm PNRD}$ in \cref{fig:restrict_PNRD}, treating any event in which one of the output photon numbers exceeds $n_{\rm PNRD}$ as a failure. As the symmetry order $N$ increases, the amplitude $\alpha$ required increases, making the success probability more strongly limited by finite detector resolution. Achieving a success probability above $99\%$ requires PNRD resolutions of $n_{\rm PNRD}=46$ and $69$ for $N=3$ and $N=4$, respectively. With post-selection, however, these requirements are reduced to $n_{\rm PNRD}=15$ for both $N=3$ and $N=4$. For this estimate, the post-selection threshold defined in \cref{sec:post-selection} is set to $\tau=0.2$, and additionally, events for which one of the output photon numbers exceeds the PNRD resolution limit are discarded.
Thus, with post-selection, the required PNRD resolution is well within the range of recent advances in PNRD technology. For example, Eaton \textit{et al.} \cite{eaton2023resolution} demonstrated photon-number resolution of up to 100 photons by multiplexing high-efficiency transition-edge sensors (TESs), and also reported resolution of up to 37 photons using a single TES channel. In addition, Cheng \textit{et al.} \cite{cheng2023100} demonstrated resolution of up to 100 photons using a 100-pixel superconducting nanowire detector.

The proposed BSM protocol based on a simple setup has several potential applications in quantum information processing. For example, Hastrup \textit{et al.} \cite{hastrup2022all} proposed an error-correction scheme for cat codes based on BSMs. Our BSM protocol could serve as a key building block for developing analogous error-correction schemes for RS-cat codes. In a related direction, Su \textit{et al.} \cite{su2022universal} proposed a universal gate set for cat codes based on gate teleportation, suggesting that similar gate-teleportation-based constructions may be possible for RS-cat codes. Beyond purely continuous-variable bosonic encodings, CV-DV hybrid approaches combining single photons with coherent states or cat codes have also been investigated \cite{lee2024fault,kiryu2025linear,lee2025photonic}. Developing methods to entangle single photons with $N$-fold RS-cat codes would therefore be an important future direction toward photonic quantum computation with improved robustness against photon loss.

\section*{Acknowledgment}
We are grateful to Tomohiro Shitara, Takahiro Kashiwazaki, and Yumiko Kasae for useful discussions. This work was supported by JST Moonshot R\&D (Grant Number JPMJMS256E).

\appendix
\counterwithin{figure}{section}
\renewcommand{\thefigure}{\Alph{section}\arabic{figure}}
\begin{widetext}
	\section{Additional proof about discrimination between \texorpdfstring{$\Phi$}{Phi} and \texorpdfstring{$\Psi$}{Psi}}
\label{appendix:additional_proof}

We show that the distribution of $X$ for the Bell basis of RS-cat codes can be approximated by a weighted sum of the distributions of $X$ for the constituent coherent states. We first consider the distribution of $X$ for
\begin{equation}
	|\xi_\delta(\alpha, \theta_1, \theta_2)\rangle \propto |\rho(\alpha, \theta_1, \delta)\rangle + |\rho(\alpha, \theta_2, -\delta)\rangle,
\end{equation}
similarly concentrates around $\cos\delta$. From \cref{eq:after_hbs}, we have
\begin{equation}
	\begin{aligned}
		U_{\mathrm{HBS}} |\rho(\alpha, \theta_1, \delta)\rangle
		= \left|\gamma_c^+\right\rangle \left|\gamma_d^+\right\rangle
		 & = e^{-\frac{|\gamma_c^+|^2 + |\gamma_d^+|^2}{2}} \sum_{n_c=0}^\infty \sum_{n_d=0}^\infty \frac{ \left(\gamma_c^+\right)^{n_c} \left(\gamma_d^+\right)^{n_d}}{\sqrt{n_c!\;n_d!}} |n_c\rangle |n_d\rangle \\
		 & = e^{-|\alpha|^2} \sum_{n_c=0}^\infty \sum_{n_d=0}^\infty \frac{ \left(\gamma_c^+\right)^{n_c} \left(\gamma_d^+\right)^{n_d}}{\sqrt{n_c!\;n_d!}} |n_c\rangle |n_d\rangle,
	\end{aligned}
\end{equation}
\begin{equation}
	\begin{aligned}
		U_{\mathrm{HBS}}|\rho(\alpha, \theta_2, -\delta)\rangle
		= \left|\gamma_c^-\right\rangle \left|\gamma_d^-\right\rangle
		 & = e^{-\frac{|\gamma_c^-|^2 + |\gamma_d^-|^2}{2}} \sum_{n_c=0}^\infty \sum_{n_d=0}^\infty \frac{ \left(\gamma_c^-\right)^{n_c} \left(\gamma_d^-\right)^{n_d}}{\sqrt{n_c!\;n_d!}} |n_c\rangle |n_d\rangle \\
		 & = e^{-|\alpha|^2} \sum_{n_c=0}^\infty \sum_{n_d=0}^\infty \frac{ \left(\gamma_c^-\right)^{n_c} \left(\gamma_d^-\right)^{n_d}}{\sqrt{n_c!\;n_d!}} |n_c\rangle |n_d\rangle,
	\end{aligned}
\end{equation}
where
\begin{equation}
	\begin{aligned}
		 & \gamma_c^+
		= \tfrac{1 + e^{i\delta}}{\sqrt2}e^{i\theta_1}\alpha
		= \sqrt2 \alpha \; \cos\frac{\delta}{2} \; e^{i(\theta_1 + \delta /2)}, \qquad
		 & \gamma_d^+
		= \tfrac{1 - e^{i\delta}}{\sqrt2}e^{i\theta_1}\alpha
		=  -i\sqrt2 \alpha \; \sin\frac{\delta}{2} \; e^{i(\theta_1 + \delta /2)}, \\
		 & \gamma_c^-
		= \tfrac{1 + e^{-i\delta}}{\sqrt2}e^{i\theta_2}\alpha
		=  \sqrt2 \alpha \; \cos\frac{\delta}{2} \; e^{i(\theta_2 - \delta /2)}, \qquad
		 & \gamma_d^-
		= \tfrac{1 - e^{-i\delta}}{\sqrt2}e^{i\theta_2}\alpha
		=  i\sqrt2 \alpha \; \sin\frac{\delta}{2} \; e^{i(\theta_2 - \delta /2)}.
	\end{aligned}
\end{equation}
Therefore,
\begin{equation}
	\begin{aligned}
		\label{eq:u_bs_xi}
		U_{\mathrm{HBS}}|\xi_\delta(\alpha, \theta_1, \theta_2)\rangle
		 & \propto U_{\mathrm{HBS}}|\rho(\alpha, \theta_1, \delta)\rangle
		+ U_{\mathrm{HBS}}|\rho(\alpha, \theta_2, -\delta)\rangle                                   \\
		 & \propto \sum_{n_c=0}^\infty \sum_{n_d=0}^\infty
		\frac{
			\left(\gamma_c^+\right)^{n_c}
			\left(\gamma_d^+\right)^{n_d}
			+ \left(\gamma_c^-\right)^{n_c}
			\left(\gamma_d^-\right)^{n_d}}
		{\sqrt{n_c!\;n_d!}}
		|n_c\rangle |n_d\rangle                                                                     \\
		 & = \sum_{n_c=0}^\infty \sum_{n_d=0}^\infty A_\delta(n_c, n_d) \; |n_c\rangle |n_d\rangle,
	\end{aligned}
\end{equation}
where we have defined
\begin{equation}
	A_\delta(n_c, n_d)
	= \frac{ \left(\gamma_c^+\right)^{n_c} \left(\gamma_d^+\right)^{n_d} + \left(\gamma_c^-\right)^{n_c} \left(\gamma_d^-\right)^{n_d}}{\sqrt{n_c!\;n_d!}}.
\end{equation}
The probability of observing the photon number $(n_c, n_d)$ is given by
\begin{equation}
	\begin{aligned}
		\label{eq:a10}
		P(n_c, n_d)
		 & \propto |A_\delta(n_c, n_d)|^2                                                                                                                     \\
		 & = \frac{|\left(\gamma_c^+\right)^{n_c} \left(\gamma_d^+\right)^{n_d} + \left(\gamma_c^-\right)^{n_c} \left(\gamma_d^-\right)^{n_d}|^2}{n_c!\;n_d!} \\
		 & = \frac{1}{n_c!\;n_d!}
		\left|(\sqrt2 \alpha \cos\tfrac{\delta}{2})^{n_c}
		(\sqrt2 \alpha \sin\tfrac{\delta}{2})^{n_d}\right|^2
		\cdot \left|e^{in_{c+d} (\theta_1 + \delta /2)} (-i)^{n_d}
		+ e^{in_{c+d} (\theta_2 - \delta/2)} \; i^{n_d}\right|^2.
	\end{aligned}
\end{equation}
By substituting $n_d = n_{c+d} - n_c$, we obtain
\begin{equation}
	\begin{aligned}
		\label{eq:a11}
		P(n_c | n_{c+d})
		 & \propto \frac{1}{n_c!\;(n_{c+d} - n_c)!}
		\left|(\sqrt2 \alpha \cos\tfrac{\delta}{2})^{n_c}
		(\sqrt2 \alpha \sin\tfrac{\delta}{2})^{n_{c+d} - n_c}\right|^2                                                                                                                                        \\
		 & \quad\quad\quad\quad \cdot \left|e^{in_{c+d} (\theta_1 + \delta /2)} (-i)^{n_{c+d} - n_c}
		+ e^{in_{c+d} (\theta_2 - \delta/2)} \; i^{n_{c+d} - n_c}\right|^2                                                                                                                                    \\
		 & \propto \frac{n_{c+d}!}{n_c!\;(n_{c+d} - n_c)!}
		\left(\cos^2\frac{\delta}{2}\right)^{n_c}
		\left(\sin^2\frac{\delta}{2}\right)^{n_{c+d} - n_c}                                                                                                                                                   \\
		 & \quad\quad\quad\quad \cdot \left[ 2 + 2 \;\mathrm{Re}\left\{e^{in_{c+d} (\theta_1 + \delta /2)} (-i)^{n_{c+d} - n_c} \; e^{-in_{c+d} (\theta_2 - \delta/2)} \; (-i)^{n_{c+d} - n_c}\right\}\right] \\
		 & \propto \frac{n_{c+d}!}{n_c!\;(n_{c+d} - n_c)!}
		\left(\cos^2\frac{\delta}{2}\right)^{n_c}
		\left(\sin^2\frac{\delta}{2}\right)^{n_{c+d} - n_c}
		\cdot \left[ 2 + 2\cdot(-1)^{n_{c+d} - n_c}\mathrm{Re}\left\{e^{in_{c+d} (\theta_1 - \theta_2 + \delta)}\right\}\right]                                                                               \\
		 & \propto \binom{n_{c+d}}{n_c}
		\left(\cos^2\frac{\delta}{2}\right)^{n_c}
		\left(\sin^2\frac{\delta}{2}\right)^{n_{c+d} - n_c}
		\cdot \left[1 + (-1)^{n_{c+d} - n_c} \cos(n_{c+d}(\theta_1 - \theta_2 + \delta))\right].
	\end{aligned}
\end{equation}
Thus, $P(n_c|n_{c+d})$ has the shape of a $\mathrm{Binomial}(n_{c+d}, \cos^2\frac{\delta}{2})$ distribution, with an additional bounded oscillatory factor:
\begin{equation}
	0 \leq 1 + (-1)^{n_{c+d} - n_c} \cos(n_{c+d}(\theta_1 - \theta_2 + \delta)) \leq 2.
\end{equation}
It follows that, when $\alpha$ is sufficiently large, the conditional distribution of $X|n_{c+d}$, and hence the distribution of $X$, concentrates around $\cos\delta$, as illustrated in \cref{fig:histogram_alpha_same}.

\begin{figure}[htbp]
	\centering
	\includegraphics[width=0.9\linewidth]{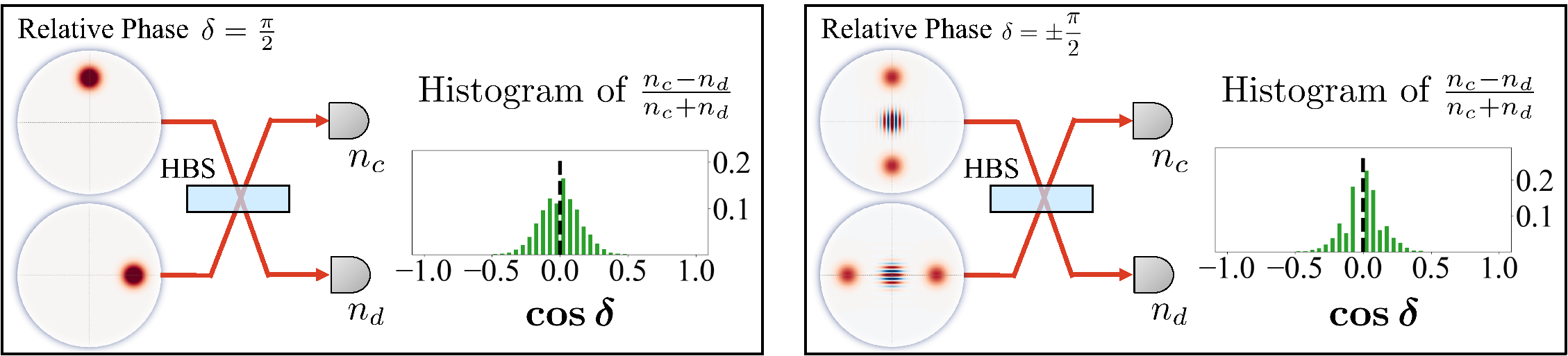}
	\caption{Histograms of $X$ with relative phases $\delta = \pm\tfrac{\pi}{2}$. The distributions are concentrated around the value indicated by the vertical dashed line, $\cos\delta=0$.}
	\label{fig:histogram_alpha_same}
\end{figure}

Next, we show that the distribution of $X$ for
\begin{equation}
	|\Xi\rangle \propto \sum_\delta w_\delta |\xi_\delta\rangle \quad (w_\delta \in \mathbb{C}),
\end{equation}
can be approximated by a linear combination of the original distributions. Using \cref{eq:u_bs_xi},
\begin{equation}
	U_{\mathrm{HBS}}|\Xi\rangle \propto \sum_{n_c=0}^\infty \sum_{n_d=0}^\infty \sum_\delta w_\delta A_\delta(n_c, n_d) \; |n_c\rangle |n_d\rangle.
\end{equation}
The probability of observing the photon number $(n_c, n_d)$ is given by:
\begin{equation}
	\tilde{P}(n_c, n_d)
	\propto \left| \sum_\delta w_\delta A_\delta(n_c, n_d) \right|^2
	= \sum_\delta |w_\delta|^2 |A_\delta(n_c, n_d)|^2 + \sum_{\delta \neq \delta'} w_\delta w^*_{\delta'} A_{\delta}(n_c, n_d)^* A_{\delta'}(n_c, n_d).
\end{equation}
The interference terms can be bounded as
\begin{equation}
	|A_{\delta}(n_c, n_d)^* A_{\delta'}(n_c, n_d)| \leq |A_{\delta}(n_c, n_d)| |A_{\delta'}(n_c, n_d)|.
\end{equation}
Since $|A_{\delta}(n_c, n_d)|$ and $|A_{\delta'}(n_c, n_d)|$ do not simultaneously take large values when $\delta \neq \delta'$,
\begin{equation}
	\tilde{P}(n_c, n_d)
	\mathrel{\underset{\sim}{\propto}} \sum_\delta |w_\delta|^2 |A_\delta(n_c, n_d)|^2 = \sum_\delta |w_\delta|^2 \; P(n_c, n_d).
\end{equation}
Consequently, the distribution of $X$ for $|\Xi\rangle = \sum_\delta  w_\delta |\xi_\delta\rangle$ can be approximated as the sum of the distributions of $|\xi_\delta\rangle$, as illustrated in \cref{fig:histogram_alpha_diff}.

\begin{figure}[htbp]
	\centering
	\includegraphics[width=0.45\linewidth]{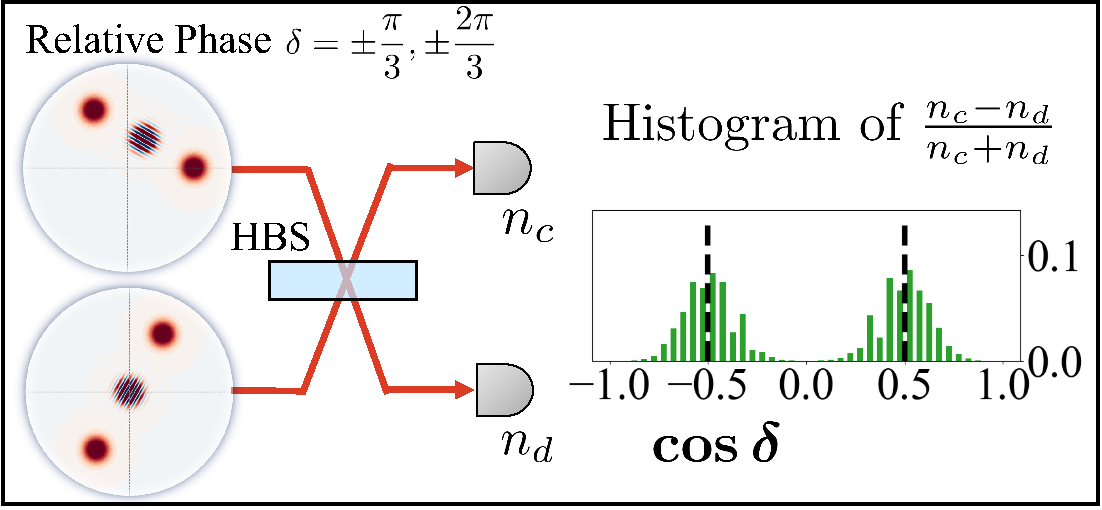}
	\caption{Histograms of $X$ with relative phases $\delta=\pm\tfrac{\pi}{3}$ and $\pm\tfrac{2\pi}{3}$. The distributions are concentrated around the values indicated by the vertical dashed lines, $\cos\delta=\pm\tfrac{1}{2}$.}
	\label{fig:histogram_alpha_diff}
\end{figure}
	\section{Overlap of the \texorpdfstring{$X$}{X} distributions of the Bell states}
\label{appendix:histogram_small_alpha}

\begin{figure}[htbp]
	\centering
	\includegraphics[width=0.8\linewidth]{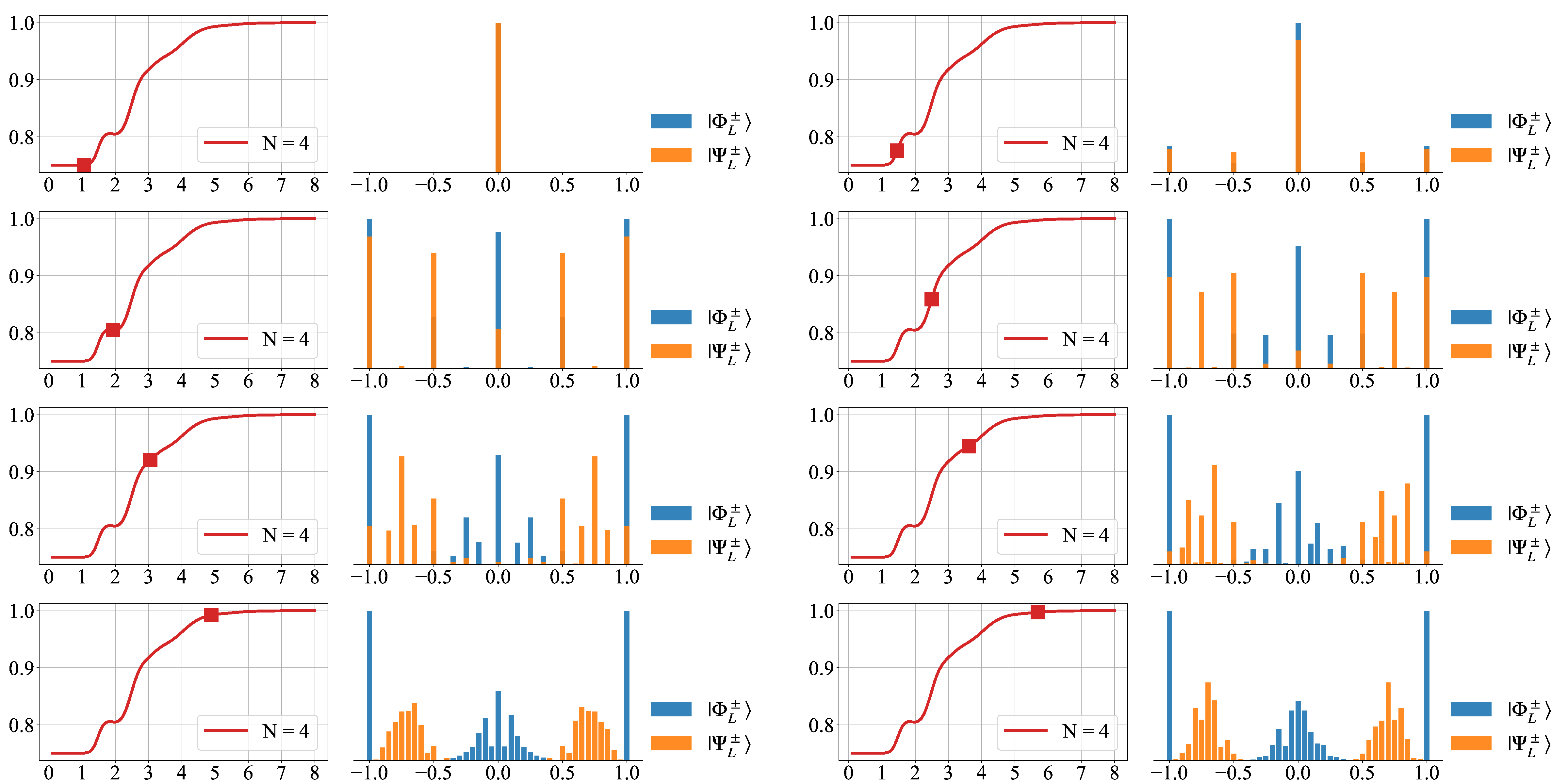}
	\caption{Histograms of $X$ for the $N=4$ RS-cat code with various amplitudes $\alpha$. The blue and orange bars correspond to $|\Phi_L^\pm\rangle$ and $|\Psi_L^\pm\rangle$, respectively. For small $\alpha$, the two distributions are not sufficiently separated, which leads to ambiguous Bell-state discrimination and limits the BSM success probability. As $\alpha$ increases, the distributions become more clearly separated.}
	\label{fig:histogram_small_alpha}
\end{figure}

In \cref{sec:bsm_rs_cat_2}, we showed that sufficiently large amplitudes $\alpha$ lead to well-separated $X$ distributions for $\{|\Phi^\pm_L\rangle\}$ and $\{|\Psi^\pm_L\rangle\}$, enabling discrimination of the Bell basis. Here, we examine the behavior of the $X$ distribution when $\alpha$ is not sufficiently large. \Cref{fig:histogram_small_alpha} shows numerically obtained histograms of $X$ for the $N=4$ RS-cat code. In this regime, the two distributions exhibit substantial overlap, making Bell-state discrimination ambiguous.

\FloatBarrier
	\section{Justification and optimality of the discriminant \texorpdfstring{$D(n_c, n_d)$}{D(nc, nd)} for Bell-state discrimination}

We introduced the discriminant $D(n_c, n_d)$ to distinguish between the sets $|\Phi^+_L\rangle$ and $|\Psi^+_L\rangle$. Here, we verify that this discriminant is optimal for discrimination based on the photon numbers $(n_c, n_d)$. \Cref{fig:discriminate_optimality} compares the success probability achieved using the discriminant $D(n_c, n_d)$ with the upper bound obtained numerically using the maximum-likelihood decision rules. The agreement between the two curves indicates that $D(n_c, n_d)$ is optimal among all discriminants that depend only on the photon numbers $(n_c, n_d)$. This optimality result applies specifically to the discrimination between $|\Phi^+_L\rangle$ and $|\Psi^+_L\rangle$. For the discrimination between $|\Phi^-_L\rangle$ and $|\Psi^-_L\rangle$, \cref{appendix:nd_parity} discusses possible improvements over the decision rule used in the main text.

\begin{figure}[htbp]
	\centering
	\begin{tikzpicture}
		\node (img) {\includegraphics[width=0.45\linewidth]{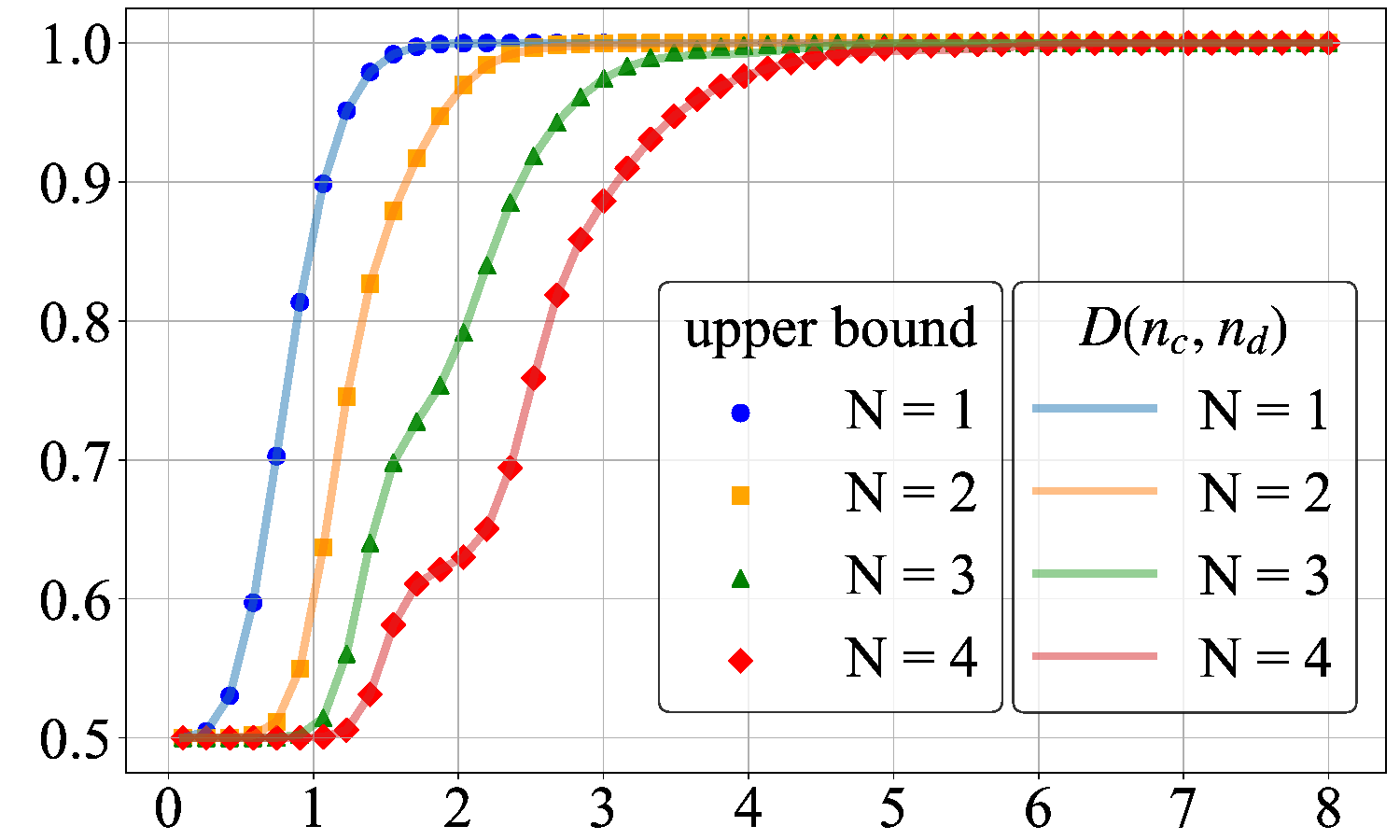}};
		\node[above=-1mm of img, xshift=1em] {Discriminate between $|\Phi_L^+\rangle$ and $|\Psi_L^+\rangle$};
		\node[below=-2mm of img, xshift=1em] {Amplitude $\alpha$};
		\node[left=1mm of img.west, rotate=90, anchor=center, xshift=1em] {Success Probability};
	\end{tikzpicture}
	\caption{Success probability for discriminating $|\Phi_L^+\rangle$ and $|\Psi_L^+\rangle$ using the discriminant $D(n_c, n_d)$, compared with the upper bound obtained by maximum-likelihood discrimination from the photon-number distribution $(n_c,n_d)$. The agreement between $D(n_c,n_d)$ and the upper bound indicates that $D(n_c,n_d)$ is optimal for photon-number-based discrimination.}
	\label{fig:discriminate_optimality}
\end{figure}
	\section{Improving the BSM success probability by considering the parity of \texorpdfstring{$n_d$}{nd}}
\label{appendix:nd_parity}

The BSM method for RS-cat codes discussed above is based on the decision rule in \cref{table:bsm_rs_cat_without_loss}. In a loss-free setting, the success probability can be further improved by taking into account the parity of $n_d$. As shown in \cref{eq:bell_fock_basis}, $|\Phi^-_L\rangle$ consists only of symmetric Fock-basis pairs of the form $|m,n\rangle + |n,m\rangle$, whereas $|\Psi^-_L\rangle$ consists only of antisymmetric Fock-basis pairs of the form $|m,n\rangle - |n,m\rangle$. When $|m,n\rangle$ is input to a HBS, the output state is given by
\begin{equation}
	U_{\mathrm{HBS}}|m, n\rangle = \frac{1}{\sqrt{m!n!}} \left(\frac{c^\dagger + d^\dagger}{\sqrt2}\right)^m
	\left(\frac{c^\dagger - d^\dagger}{\sqrt2}\right)^n |0, 0\rangle.
\end{equation}
The probability amplitude for detecting $n_d = r$ photons in the $d$ port is therefore
\begin{equation}
	A_{m+n-r, r} = 2^{-\frac{m+n}{2}}\sum_{p=0}^r \binom{m}{r-p} \binom{n}{p} (-1)^p.
\end{equation}
Similarly, when $|n,m\rangle$ is input to the HBS, the output state is
\begin{equation}
	U_{\mathrm{HBS}}|n, m\rangle = \frac{1}{\sqrt{m!n!}} \left(\frac{c^\dagger + d^\dagger}{\sqrt2}\right)^n
	\left(\frac{c^\dagger - d^\dagger}{\sqrt2}\right)^m |0, 0\rangle,
\end{equation}
and the corresponding probability amplitude is
\begin{equation}
	B_{m+n-r, r}
		= 2^{-\frac{m+n}{2}}\sum_{p=0}^r \binom{m}{r-p} \binom{n}{p} (-1)^{r-p}
	= 2^{-\frac{m+n}{2}}\sum_{p=0}^r \binom{m}{r-p} \binom{n}{p} (-1)^{r+p}
	= (-1)^r A_{m+n-r, r}.
\end{equation}
Consequently,
\begin{equation}
	A_{m+n-r, r} \pm B_{m+n-r, r} = \left[1 \pm (-1)^r\right] A_{m+n-r, r}.
\end{equation}
Thus, when a symmetric Fock-basis pair $|m,n\rangle + |n,m\rangle$ passes through a HBS, the photon number in the $d$ port is even, whereas for an antisymmetric Fock-basis pair $|m,n\rangle - |n,m\rangle$, the photon number in the $d$ port is odd. Therefore, $|\Phi^-_L\rangle$ and $|\Psi^-_L\rangle$ can be distinguished by the parity of the photon number in the $d$ port.

Based on this observation, the BSM can be performed using the decision rule shown in \cref{fig:bsm_rs_cat_improved}(a). Furthermore, as shown in \cref{fig:bsm_rs_cat_improved}(b), the success probability can be further improved in a loss-free environment by incorporating the parity of $n_d$. This improved success probability agrees with the upper bound given by the maximum-likelihood decision rule. In the presence of photon loss, however, the parity of $n_d$ is readily flipped, making this approach less practical.

\begin{figure}[htbp]
	\centering
	\begin{minipage}{0.4\textwidth}
		\centering
		\textbf{(a)}\par\vspace{1mm}

		\begin{tabular}{ccc}
			\toprule
			\multicolumn{2}{c}{Measurement Pattern}
			 & Decision                        \\
			\cmidrule(r){1-2} \cmidrule(l){3-3}
			\multirow{2}{*}{$n_c + n_d \equiv 0 \pmod{2N}$}
			 & $D(n_c, n_d) \equiv 0 \pmod{2}$
			 & $|\Phi^+_L\rangle$              \\
			 & $D(n_c, n_d) \equiv 1 \pmod{2}$
			 & $|\Psi^+_L\rangle$              \\
			\midrule
			\multirow{2}{*}{$n_c + n_d \equiv N \pmod{2N}$}
			 & $n_d \equiv 0 \pmod{2}$
			 & $|\Phi^-_L\rangle$              \\
			 & $n_d \equiv 1 \pmod{2}$
			 & $|\Psi^-_L\rangle$              \\
			\bottomrule
		\end{tabular}
	\end{minipage}
	\hspace{3em}
	\begin{minipage}{0.5\textwidth}
		\centering
		\textbf{(b)}\par\vspace{1mm}
		\begin{tikzpicture}
			\node (img) {\includegraphics[width=0.93\linewidth]{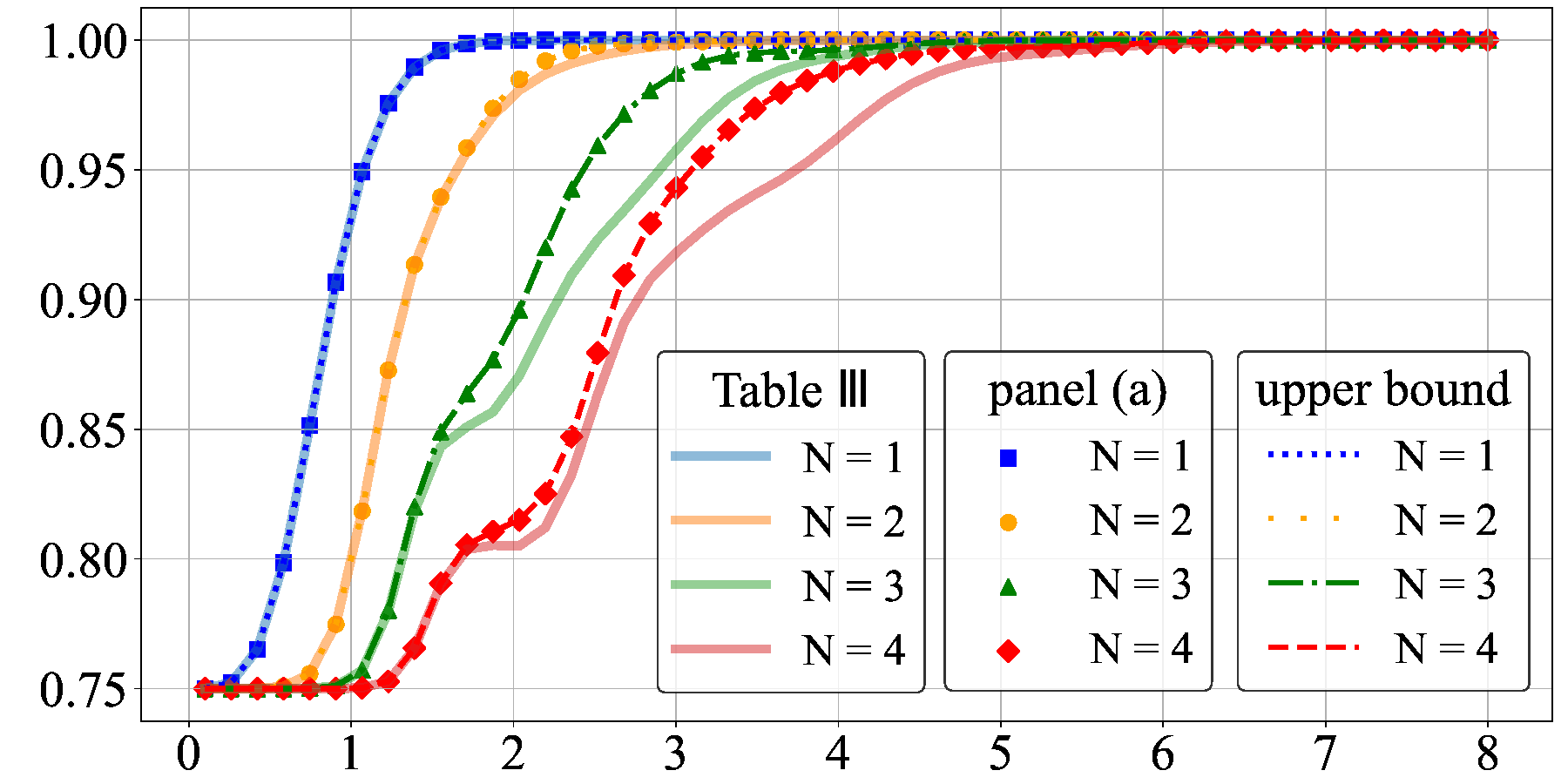}};
			\node[below=-2mm of img, xshift=1em] {Amplitude $\alpha$};
			\node[left=-1mm of img.west, rotate=90, anchor=center, xshift=1em] {Success Probability};
		\end{tikzpicture}
	\end{minipage}
	\caption{Success probabilities of the BSM for RS-cat codes using the parity of $n_d$. (a) Decision rule for the BSM using the parity of $n_d$. (b) Comparison of the success probabilities obtained from \cref{table:bsm_rs_cat_without_loss} and panel (a) for symmetry orders $N=1,2,3,$ and $4$.}
	\label{fig:bsm_rs_cat_improved}
\end{figure}

	\section{Histogram of \texorpdfstring{$n_c - n_d$}{nc - nd} and BSM success probability}

The BSM method introduced above relies on the normalized measurement outcome $X$, whose distribution is concentrated around $\cos\delta$ when a pair of coherent states $|\rho(\alpha, \theta, \delta)\rangle = |e^{i\theta}\alpha\rangle |e^{i(\theta + \delta)}\alpha\rangle$ is injected into a HBS. More simply, Bell states can also be identified from the distribution of
\begin{equation}
	\tilde{X} = n_c - n_d.
\end{equation}
As shown in \cref{eq:nc_nd_poisson}, $n_c$ and $n_d$ each independently follow a Poisson distribution. Therefore, $\tilde{X} = n_c - n_d$ follows a Skellam distribution given by
\begin{equation}
	P(\tilde{X} = k) = e^{-(\mu_c + \mu_d)} \left(\frac{\mu_c}{\mu_d} \right)^{k/2} J_{|k|}(2\sqrt{\mu_c\mu_d}),
\end{equation}
where $J_{|k|}(x)$ is the modified Bessel function of the first kind:
\begin{equation}
	J_{|k|}(x) = \sum_{m=0}^\infty \frac{1}{m!\;(m+|k|)!} \left(\frac{x}{2}\right)^{2m+|k|}.
\end{equation}
This distribution, as illustrated in \cref{fig:histogram_rho_dash}, has a mean and variance of
\begin{equation}
	E[\tilde{X}] = \mu_c - \mu_d = 2\alpha^2\cos\delta, \qquad \mathrm{Var}(\tilde{X}) = \mu_c + \mu_d = 2\alpha^2.
\end{equation}

\begin{figure}[htbp]
	\centering
	\includegraphics[width=0.5\linewidth]{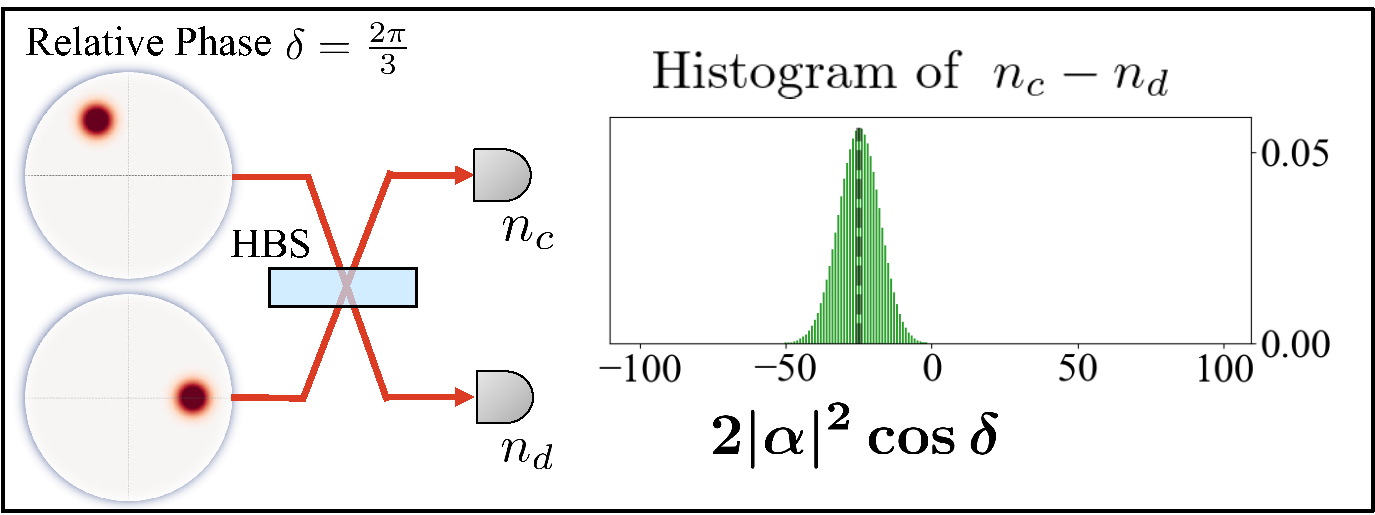}
	\caption{Histogram of $\tilde{X}=n_c-n_d$ for two coherent states $|\alpha\rangle |e^{2\pi i/3}\alpha\rangle$ with relative phase $\delta=2\pi/3$. The histogram is concentrated around the value indicated by the vertical dashed line, $2\alpha^2\cos\delta$.}
	\label{fig:histogram_rho_dash}
\end{figure}

\begin{figure}[htbp]
	\centering
	\includegraphics[width=0.8\linewidth]{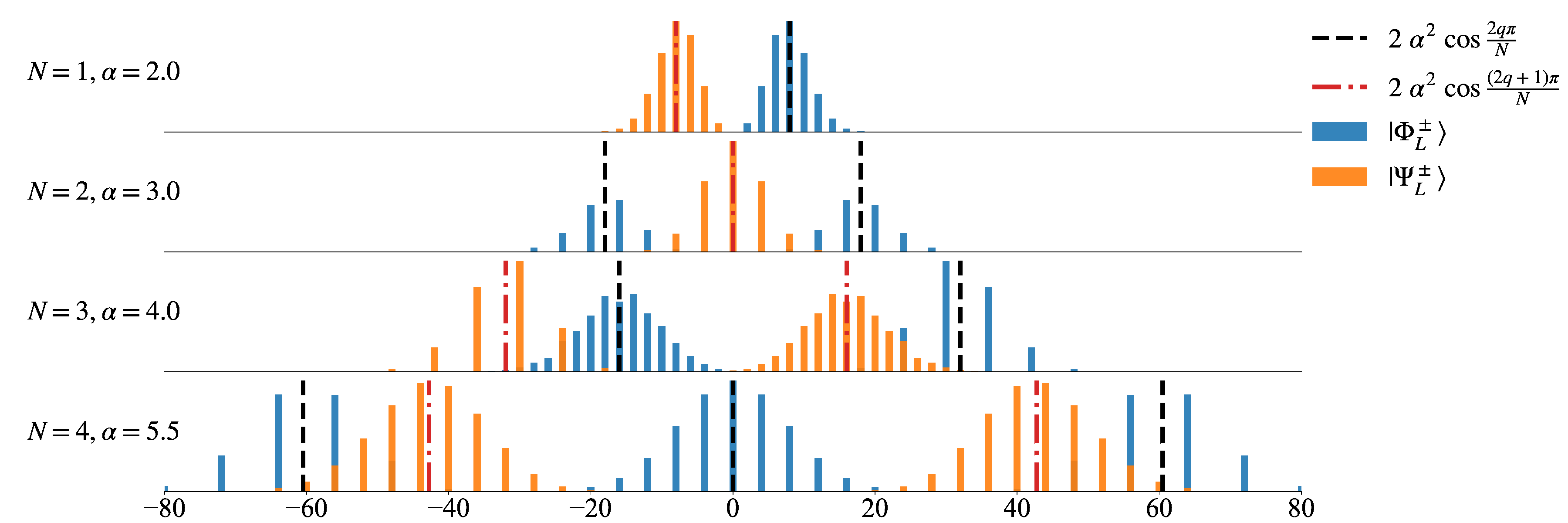}
	\caption{Histograms of $\tilde{X}=n_c-n_d$ for the Bell basis states of the RS-cat code. The blue and orange bars correspond to $|\Phi_L^\pm\rangle$ and $|\Psi_L^\pm\rangle$, respectively. The histograms are concentrated around the values indicated by the vertical dashed lines, with the black lines denoting ${2\alpha^2\cos(2q\pi/N)}{q=0}^{N-1}$ and the red lines denoting ${2\alpha^2\cos((2q+1)\pi/N)}{q=0}^{N-1}$.}
	\label{fig:histogram_cat_nc_minus_nd}
\end{figure}

As shown in \cref{fig:histogram_cat_nc_minus_nd}, the distribution of $\tilde{X}$ for the Bell basis of RS-cat codes can be approximated by a superposition of the corresponding coherent-state distributions. Each component has mean and variance $E[\tilde{X}] = 2\alpha^2\cos\delta, \quad \mathrm{Var}(\tilde{X}) = 2\alpha^2$. Since the variance is independent of $\delta$, the widths of these distributions remain finite even near $\delta=0$ and $\delta=\pi$, where the separation between neighboring mean values becomes small.

By contrast, the normalized measurement outcome $X$ has component distributions with $E[X] = \cos\delta, \quad \mathrm{Var}(X) \simeq \sin^2\delta / (2\alpha^2)$. In this case, the variance decreases near $\delta=0$ and $\delta=\pi$, which suppresses the overlap between neighboring distributions. Consequently, the BSM based on $\tilde{X}=n_c-n_d$ has a lower success probability than the BSM based on the decision rule in \cref{table:bsm_rs_cat_without_loss}. The corresponding success probabilities are compared in \cref{fig:success_rate_nc_minus_nd}.

\begin{figure}[htbp]
	\centering
	\begin{tikzpicture}
		\node (img) {\includegraphics[width=0.45\linewidth]{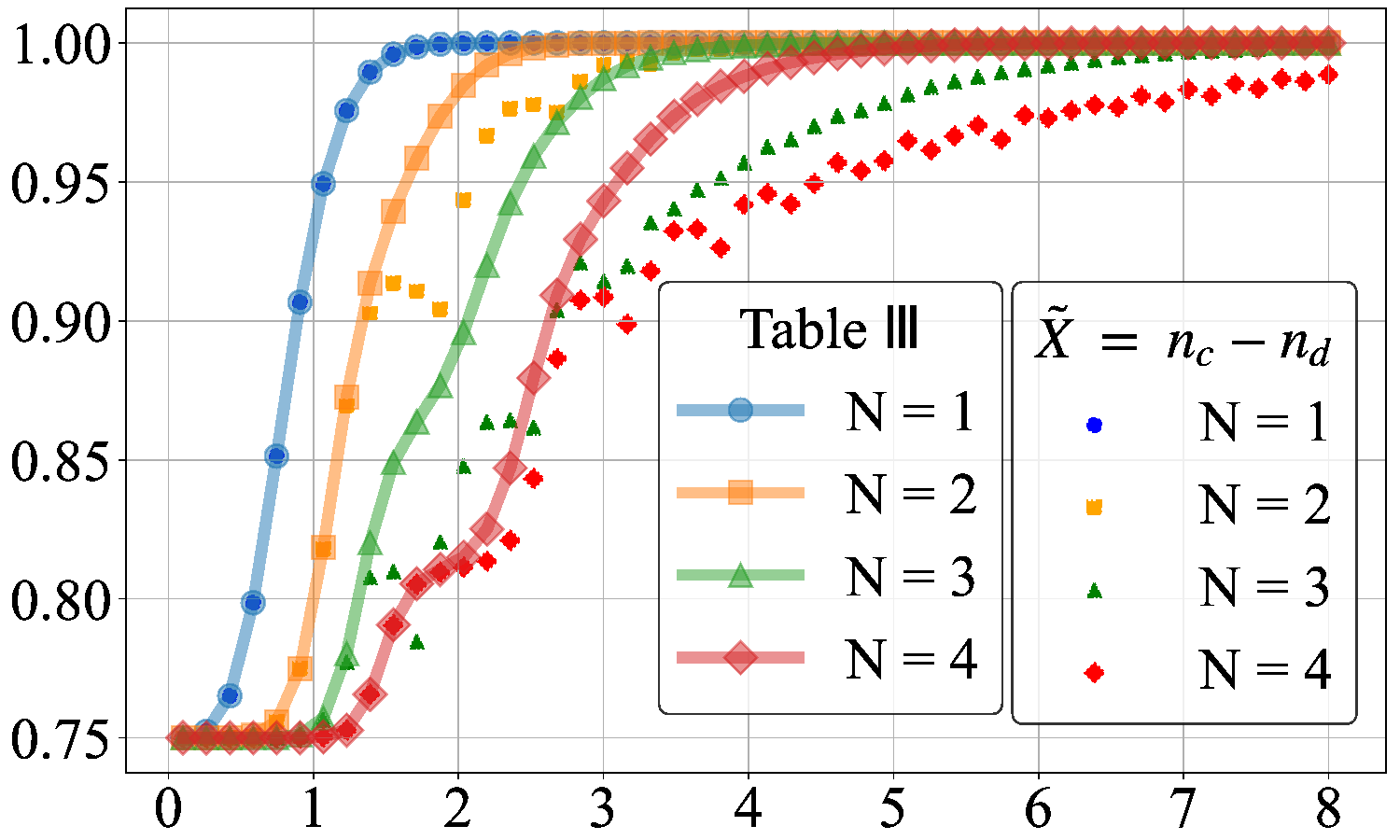}};
		\node[below=-2mm of img, xshift=1em] {Amplitude $\alpha$};
		\node[left=1mm of img.west, rotate=90, anchor=center, xshift=1em] {Success Probability};
	\end{tikzpicture}
	\caption{Comparison of the BSM success probabilities for RS-cat codes under lossless conditions. The solid lines correspond to the decision rule in \cref{table:bsm_rs_cat_without_loss}, whereas the dashed lines correspond to the discrimination based on $\tilde{X}=n_c-n_d$. The success probabilities are shown as functions of the amplitude $\alpha$ for $N=1,2,3$, and $4$.}
	\label{fig:success_rate_nc_minus_nd}
\end{figure}

\FloatBarrier
	\section{BSM for RS-binomial codes}
\label{sec:binomial}

\begin{figure}[htbp]
	\centering
	\begin{minipage}{0.2\textwidth}
		\centering
		$N=1, K=10$
		\includegraphics[width=\textwidth]{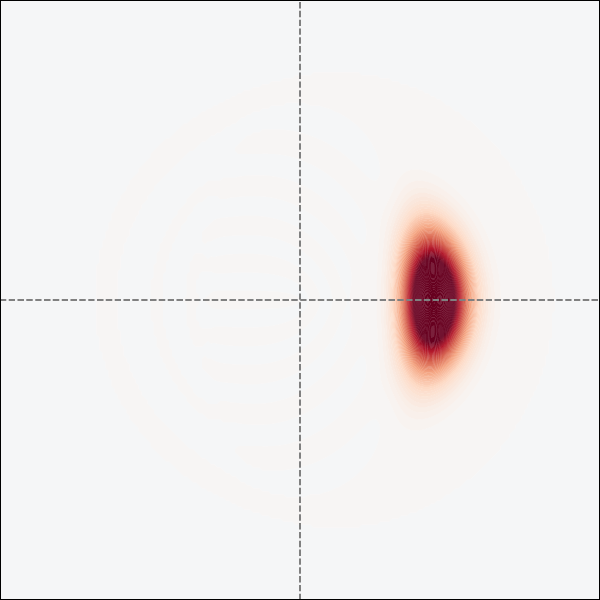}
	\end{minipage}
	\hspace{10pt}
	\begin{minipage}{0.2\textwidth}
		\centering
		$N=2, K=20$
		\includegraphics[width=\textwidth]{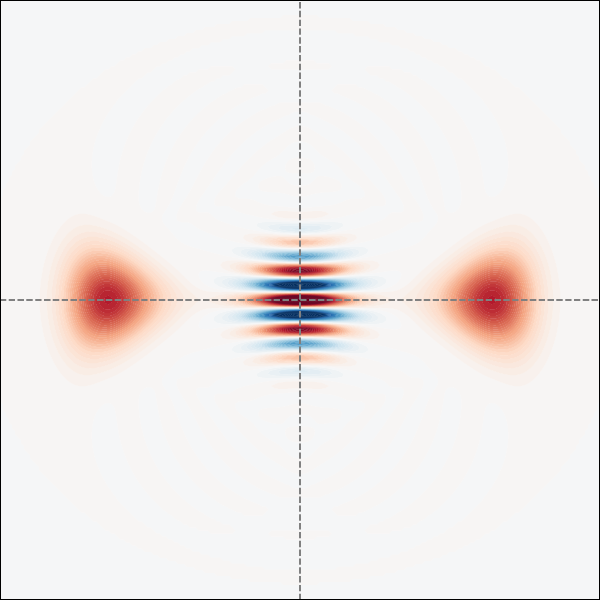}
	\end{minipage}
	\hspace{10pt}
	\begin{minipage}{0.2\textwidth}
		\centering
		$N=3, K= 30$
		\includegraphics[width=\textwidth]{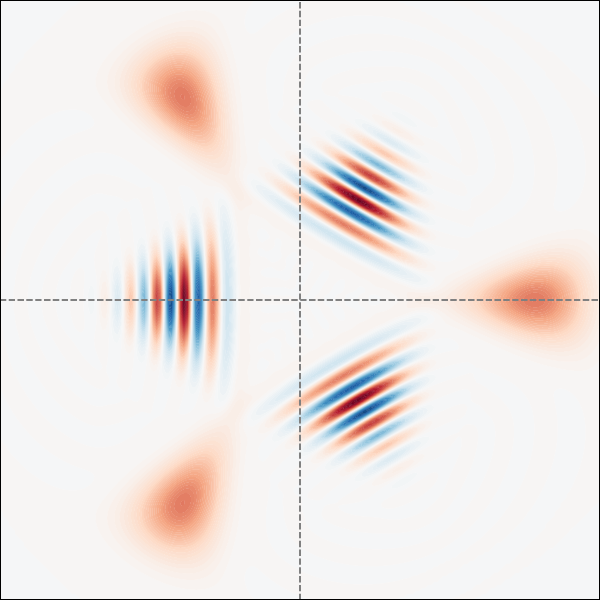}
	\end{minipage}
	\hspace{10pt}
	\begin{minipage}{0.2\textwidth}
		\centering
		$N=4, K = 40$
		\includegraphics[width=\textwidth]{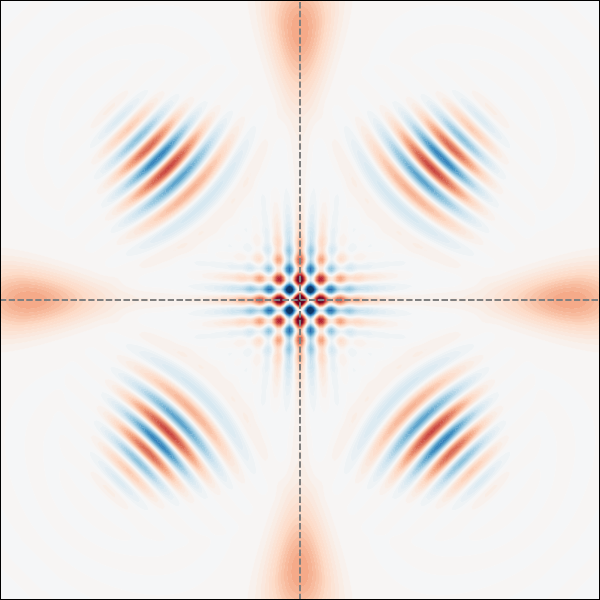}
	\end{minipage}
	\caption{Wigner functions of RS-binomial codes. The panels show the cases $N = 1, 2, 3$, and $4$, with the corresponding cutoff parameters $K = 10, 20, 30$, and $40$ indicated above each plot.}
	\label{fig:wigner_function_binomial}
\end{figure}

One of the codes possessing rotational symmetry, similar to the RS-cat codes, are the rotation-symmetric binomial (RS-binomial) codes. For a given cutoff parameter $K \in \mathbb{N}$, the RS-binomial codes are defined using the Fock basis as follows:
\begin{equation}
	|0_L\rangle = \mathcal{N}_0 \sum_{k=0}^{K} \sqrt{\frac{1}{2^{K-1}}\binom{K}{k}} |kN\rangle, \quad
	|1_L\rangle  = \mathcal{N}_0 \sum_{k=0}^{K} (-1)^k \sqrt{\frac{1}{2^{K-1}}\binom{K}{k}} |kN\rangle.
\end{equation}
The RS-binomial codes possess a photon number distribution similar to that of RS-cat codes. The Wigner functions of the RS-binomial codes for $N=1, 2, 3, 4$ are shown in \cref{fig:wigner_function_binomial}. Here, we investigate the success probability of the BSM when the proposed method is applied to the RS-binomial codes. As with the case of RS-cat codes, we evaluate the success probability for RS-binomial codes with $N=1, 2, 3$, and $4$.

\begin{figure}[htbp]
	\centering
	\begin{tikzpicture}
		\node (img) {\includegraphics[width=0.45\linewidth]{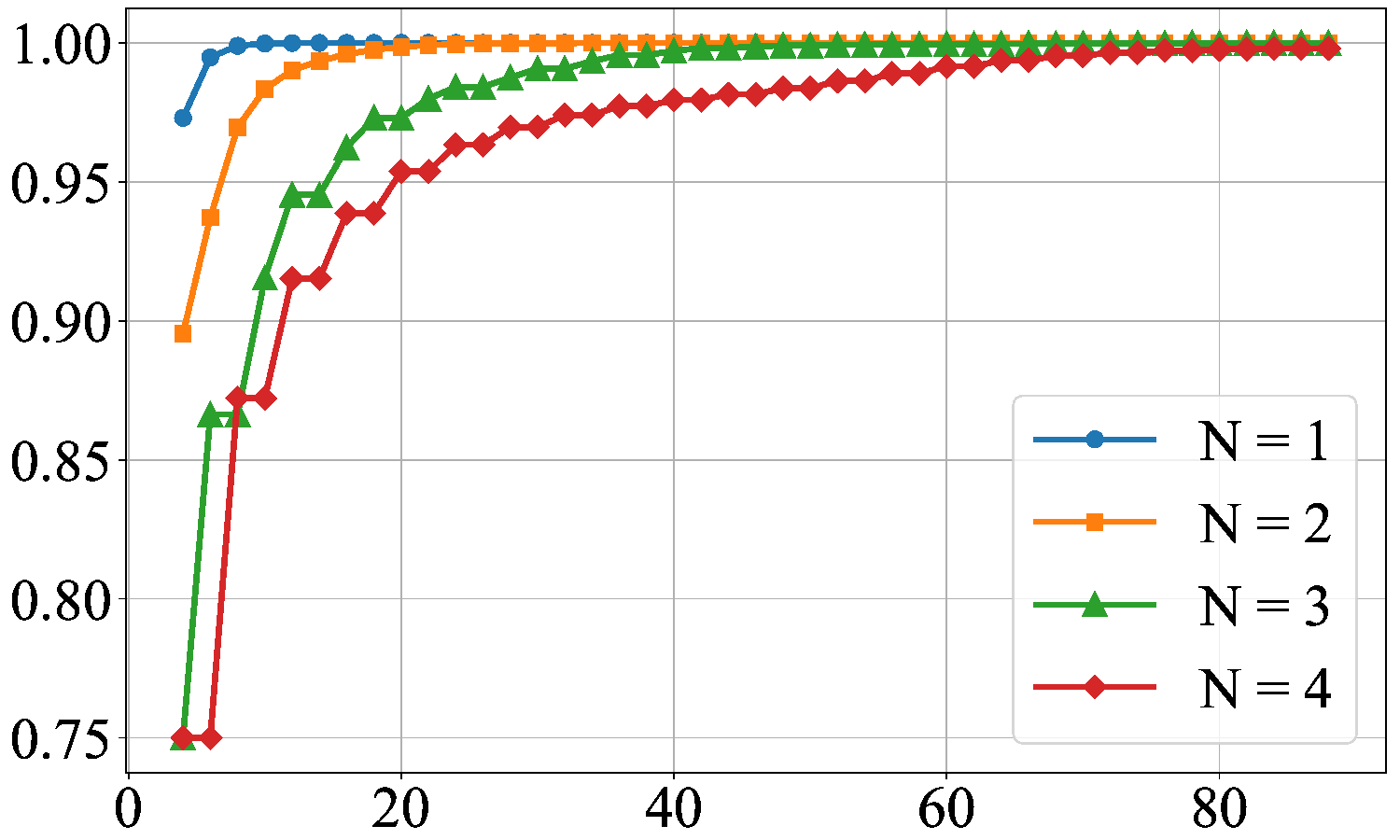}};
		\node[below=-2mm of img, xshift=1em] {Cutoff $K$};
		\node[left=1mm of img.west, rotate=90, anchor=center, xshift=1em] {Success Probability};
	\end{tikzpicture}
	\caption{Success probability of the BSM for RS-binomial codes. The horizontal axis represents the cutoff parameter $K$, and the vertical axis represents the BSM success probability. The blue, orange, green, and red curves correspond to symmetry orders $N=1,2,3$, and $4$, respectively. For each symmetry order $N$, the BSM becomes deterministic as the cutoff parameter $K$ increases.}
	\label{fig:binomial}
\end{figure}

The simulation results are shown in \cref{fig:binomial}. The horizontal axis represents the magnitude of the cutoff parameter $K$, and the vertical axis indicates the success probability of the BSM. These results demonstrate that near-deterministic BSM can be achieved for RS-binomial codes with a sufficiently large cutoff parameter $K$. Furthermore, similar to the case of RS-cat codes, as the symmetry order $N$ increases, a larger cutoff parameter $K$ is required to perform deterministic BSM.

\FloatBarrier
\end{widetext}

\bibliography{reference}

\end{document}